\documentclass[12pt]{extarticle}
\usepackage[utf8]{inputenc}
\usepackage[margin=1in]{geometry}
\usepackage{mathtools}
\usepackage{booktabs}
\usepackage{amsmath}
\usepackage{appendix}
\usepackage{graphicx}
\usepackage[para]{threeparttable}
\usepackage[round]{natbib}
\usepackage{setspace}
\usepackage{lineno}
\usepackage{lipsum}
\usepackage{url}
\usepackage{hyperref}
\usepackage{adjustbox}
\usepackage{float}
\usepackage{tikz}
\usepackage{amsmath}
\usepackage{mathabx}
\usepackage{hyperref}
\usepackage{multirow}
\usepackage{listings}

\usetikzlibrary{decorations.pathreplacing}
\usetikzlibrary{arrows}

\begin{document}

\title{\Large{\textbf{An Augmented Likelihood Approach for the Discrete Proportional Hazards Model Using Auxiliary and Validated Outcome Data – with Application to the HCHS/SOL Study}}}
\author{Lillian A. Boe$^{1,*}$ and 
Pamela A. Shaw$^{2}$ \\
$^{1}$Department of Biostatistics, Epidemiology, and Informatics, \\ University of Pennsylvania Perelman School of Medicine, \\  Philadelphia, PA 19104\\
$^{2}$Biostatistics Unit, Kaiser Permanente Washington Health Research Institute, \\ Seattle, Washington 98101 \\
\textit{*email}: boel@pennmedicine.upenn.edu}

\date{}
\maketitle

\begin{abstract}
{In large epidemiologic studies, it is typical for an inexpensive, non-invasive procedure to be used to record disease status during regular follow-up visits, with less frequent assessment by a gold standard test. Inexpensive outcome measures like self-reported disease status are practical to obtain, but can be error-prone. Association analysis reliant on error-prone outcomes may lead to biased results; however, restricting analyses to only data from the less frequently observed error-free outcome could be inefficient. We have developed an augmented likelihood that incorporates data from both error-prone outcomes and a gold standard assessment. We conduct a numerical study to show how we can improve statistical efficiency by using the proposed method over standard approaches for interval-censored survival data that do not leverage auxiliary data. We extend this method for the complex survey design setting so that it can be applied in our motivating data example. Our method is applied to data from the Hispanic Community Health Study/Study of Latinos to assess the association between energy and protein intake and the risk of incident diabetes. In our application, we demonstrate how our method can be used in combination with regression calibration to additionally address the covariate measurement error in self-reported diet.}
\end{abstract}

Measurement error, misclassification, surrogate endpoint, augmented likelihood, proportional hazards, survival analysis.

\maketitle

\section{Introduction}
In large epidemiologic or clinical studies with periodic follow-up, it is often impractical to obtain a gold standard or reference standard test on all subjects at each visit time throughout the study. Instead, an inexpensive measures is typically used to assess the outcome of interest at each follow-up visit , and the reference standard diagnostic test is obtained less frequently, if at all.  Compared to some reference standard diagnostic tests that may involve invasive or otherwise impractical biomarkers, self-reported disease status is inexpensive, noninvasive, and relatively easy to obtain in large cohorts. However, self-reported disease status is often prone to measurement error. For example, some studies have shown that the sensitivity and specificity of self-reported diabetes are imperfect compared to the reference instruments of fasting glucose and hemoglobin A1c (HbA1c).\citep{gu2015semiparametric,margolis2008validity}

There has been considerable interest in methods that use surrogate or auxiliary data to improve the efficiency of inference for time-to-event analyses. In this context, surrogate endpoints are defined as outcomes that are intended to replace the true, or gold standard, outcome of interest, while auxiliary data refers to variables that are used to improve the efficiency of the analysis of the gold standard endpoint.\citep{conlon2015improving}
Pepe (1992)\citep{pepe1992inference} introduced an estimated likelihood method for general data structures in which surrogate outcomes are available on all subjects and true outcomes are available on a subset. Magaret (2008)\citep{magaret2008incorporating} extended this work to the setting of the discrete proportional hazards model. Zee et al. (2015)\citep{zee2015assessing} proposed a similar semiparametric estimated likelihood approach for parameter estimation that allows for real-time validation and does not require true and surrogate censoring times to be equal when the surrogate outcome is censored.  Fleming et al. (1994) \citep{fleming1994surrogate} presented an augmented likelihood approach that incorporates auxiliary information into the proportional hazards model for cases when true endpoints are available on all study subjects. In their method, the likelihood can be augmented for subjects using an auxiliary (surrogate) outcome whose true endpoints are censored prior to their auxiliary endpoints.  

Several methods have been developed to correct errors in binary outcome variables for discrete time-to-event settings when gold or reference standard outcome data are not available. For these approaches, estimated values of sensitivity and specificity are incorporated into the analysis to correct for the bias induced by errors in the event classification variable. Specifically, Meier et al. (2003) \citep{meier2003discrete} introduced an adjusted proportional hazards model for estimating hazard ratios in the presence of discrete failure time data subject to misclassification. \cite{gu2015semiparametric} developed a likelihood-based method that models the association of a covariate with a discrete time-to-event outcome recorded by error-prone self-reports or imperfect diagnostic tests, assuming the proportional hazards model. \cite{boe2021approximate} extended this work by incorporating regression calibration to additionally adjust for covariate measurement error for cases in which one or more exposure variables of interest are also recorded with error. Each of these methods addressed the misclassification by incorporating externally estimated sensitivity and specificity into the estimation.

In this paper, we develop an augmented likelihood approach that incorporates error-prone auxiliary data into the analysis of an interval-censored, gold standard assessment of a time-to-event outcome. Our method is distinct from prior work in that we consider the setting where subjects have both frequent follow-up with an auxiliary outcome and infrequent follow-up with a gold standard evaluation. Our method may be applied when auxiliary outcome data, observed through periodically collected self-reports or diagnostic tests, are available either before or after the gold standard is scheduled to be observed.  This work is motivated by the Hispanic Community Health Study/Study of Latinos (HCHS/SOL), a prospective longitudinal cohort with (1) a reference standard biomarker-defined diabetes status variable, using fasting glucose and/or hemoglobin A1c (HbA1c), available at baseline and once more after 4-10 years, and (2) self-reported diabetes status recorded annually, up to 4 years beyond the reference test.

 We begin the next section by introducing notation and presenting the theoretical development of our augmented likelihood function. We also extend our method to handle data from a complex survey design and develop a sandwich variance estimator. In section \ref{section3}, we provide an extensive numerical study to demonstrate how we can improve statistical efficiency by using the proposed method instead of standard approaches for interval-censored survival data that do not leverage the auxiliary data. Section \ref{section4} introduces the HCHS/SOL study and illustrates the results of applying the proposed approach to this data set to assess the association between dietary energy, protein, and protein density intake and incident diabetes. For this analysis, we additionally address the covariate measurement error. We conclude by providing a discussion of our findings and potential extensions of this work in section \ref{discussion}. 
\section{Methods}\label{allmethods}
\subsection{Notation and Time-to-Event Model}\label{ref_section1} 
Define $T_i$ as the unobserved, continuous event time  of interest for subjects $i=1,...,N$. We assume the setting of a prospective cohort study where the participants follow-up occurs at regular visit intervals (e.g. annually) and all subjects are known to be disease-free at baseline, time $\tau_0$. Let $0=\tau_0<\tau_1 < \ldots \tau_J$ be the possible visit times among the $N$ subjects and $\tau_{J+1}=\infty$. Thus, the possible follow-up can be broken into  $J+1$ disjoint intervals as follows: $[\tau_0,\tau_1),[\tau_1,\tau_2),...[\tau_J,\tau_{J+1})$. Let $n_i$ be the number of visits for the $i^{th}$ subject, which we assume is random.  We consider the setting where each subject reports a potentially error-prone disease status at each visit   until the first positive self-report or censoring time. Let $\mathbf{Y^*_i}$ be the vector of error-prone binary outcomes that indicates whether the $i$th subject self-reported the event at time $j$, and $\mathbf{T^*_i}$ be the corresponding vector of visit times. More specifically, we define $Y_{ij}^*$ as the binary indicator that the $j^{th}$ self-report for the $i^{th}$ subject is positive. Motivated by the design of the HCHS/SOL study, we assume that a gold standard assessment of the disease is also obtained, but only once post-baseline. Namely, we assume at time $\tau_{V_i}$, we observe $\Delta_i$, a scalar binary indicator for each subject's true disease status recorded by a gold standard diagnostic test, where $V_i \in \{1,2,...,J\}$. Note that in some studies, the time of gold standard assessment is fixed at $\tau_{V_i}=\tau_{J}$ for all subjects, but we allow $\tau_{V_i} \leq \tau_{J}$, suggesting that follow-up by self-report may continue after the gold standard outcome is reported. Finally, we assume that as a result of loss to follow-up, $\Delta_i$ may be missing on a subset of study subjects and define $M_i$ as the binary variable indicating whether $\Delta_i$ is missing. We assume this outcome is missing completely at random, though missing at random patterns could also be readily incorporated with application of a standard inverse probability weighting approach. We can now write the joint probability of the observed data for the $i^{th}$ subject as $P(\mathbf{Y^*_i}, \mathbf{T^*_i}, \Delta_i, T_i) = P(\mathbf{Y^*_i}, \mathbf{T^*_i}| \Delta_i, T_i )P(\Delta_i, T_i) = \sum_{j=1}^{J+1} P(\mathbf{Y^*_i}, \mathbf{T^*_i}| \Delta_i, \tau_{j-1}<T_i \leq \tau_j)P(\Delta_i, \tau_{j-1}<T_i \leq \tau_j).$

Following previous work to address misclassified outcomes in the discrete proportional hazards model, \cite{boe2021approximate,gu2015semiparametric,balasubramanian2003estimation} we assume the $n_i$ error-prone outcomes $Y^*_{ij}$ are conditionally independent given the true disease status and event time $T_i$, such that $P(\mathbf{Y^*_i}|T_i,\mathbf{T_i^*},\Delta_i)=\prod_{l=1}^{n_i}P(Y^*_{il}|T_i,T_{il}^*,\Delta_i)$. We can re-express the joint probability of observed data for the $i$th subject as follows:
\begin{eqnarray}\label{likelihoodnocov}
P(\mathbf{Y^*_i}, \mathbf{T_i^*}, \Delta_i, T_i) 
&=& \sum_{j=1}^{J+1}  C_{ij} P(\Delta_i, \tau_{j-1}<T_i \leq \tau_j)
,
\end{eqnarray}

\noindent where $C_{ij}=\left[\prod_{l=1}^{n_i}P(Y^*_{il}|\tau_{j-1}<T_i\leq \tau_j,T_{il}^*,\Delta_i)\right].$    We first assume that sensitivity $(Se)$ and specificity $(Sp)$ are known constants and have the following definitions: $Se=\Pr(Y^*_{il}=1|\tau_{j-1}<T_i\leq\tau_j,T_l^*\geq\tau_j)$ and $Sp=\Pr(Y^*_{il}=0|\tau_{j-1}<T_i\leq\tau_j,T_l^*\leq\tau_{j-1})$. Then, the $C_{ij}$ are simply functions of the sensitivity and specificity. See section S2 of the Supplementary Materials for details.

We will now derive the likelihood contribution for subjects with observed $\Delta_i$ (i.e., $M_i=0$). For these subjects, we can rewrite the likelihood in equation \ref{likelihoodnocov} as follows:
\begin{eqnarray}\label{likelihoodnocovdelta1}
P(\mathbf{Y^*_i}, \mathbf{T_i^*}, \Delta_i, T_i) 
&=&  \sum_{j=1}^{J+1}  C_{ij} P(\tau_{j-1}<T_i\leq \tau_j|\Delta_i) P(\Delta_i).
\end{eqnarray}
Define $\theta_j=\Pr(\tau_{j-1}<T_i\leq\tau_j)$. If at time $\tau_{V_i}$, subject $i$ is identified as a validated positive, then we have $P(\Delta _i=1)=P(T_i \leq \tau_{V_i})=\sum_{l=1}^{V_i} \theta_l$ and:
\begin{eqnarray*}
P(\tau_{j-1}<T_i\leq \tau_j|\Delta_i=1) &=&\begin{cases} \frac{\theta_j}{\sum_{l=1}^{V_i} \theta_l} &\mbox{for } 1\leq j \leq V_i \\
0 & \mbox{for } V_i < j \leq J+1. \end{cases}  
\end{eqnarray*}

If subject $i$ is identified to be a validated negative at time $\tau_{V_i}$, then $P(\Delta _i=0)=P(T_i > \tau_{V_i})=\sum_{l=V_i+1}^{J+1} \theta_l$ and: 
\begin{eqnarray*}
P(\tau_{j-1}<T_i\leq \tau_j|\Delta_i=0) &=&\begin{cases} 0 &\mbox{for } 1\leq j \leq V_i \\
\frac{\theta_j}{\sum_{l=V_i+1}^{J+1} \theta_l} & \mbox{for } V_i < j \leq J+1. \end{cases}
\end{eqnarray*}

Next, we derive the likelihood for a subject who is lost to follow-up and is missing $\Delta_i$ (i.e. $M_i=1$). In this scenario, the joint probability of observed data for the $i$th subject is $P(\mathbf{Y^*_i}, \mathbf{T_i^*}, \Delta_i, T_i) 
= \sum_{j=1}^{J+1} \Bigg[ \prod_{l=1}^{n_i}P(Y^*_{il}|\tau_{j-1}<T_i\leq \tau_j,T_{il}^*) \Bigg] P(\tau_{j-1}<T_i\leq \tau_j)$, and thus:
\begin{eqnarray}\label{likelihoodnocovdelta0}
P(\mathbf{Y^*_i}, \mathbf{T_i^*}, \Delta_i, T_i)  &=& \sum_{j=1}^{J+1}  C_{ij} \theta_j.
\end{eqnarray}
Define $X_i$ as the $p$-dimensional vector of time-invariant covariates. We assume that $X$ is related with the outcome through a Cox proportional hazards model, $S(t)=S_0(t)^{\exp(x'\beta)}$. We use this model to re-express the joint probability from equations \ref{likelihoodnocovdelta1} and \ref{likelihoodnocovdelta0} and write the likelihood in terms of the baseline survival probabilities, $\mathbf{S}=(S_1,S_2,...,S_{J+1})'$, where $S_j=\Pr(T_0>\tau_{j-1})$ and $T_0$ is a random variable that has survival function $S_0(t)$. Thus $1=S_1>S_2>...>S_{J+1}>0$ and $S_j=\sum_{h=j}^{J+1}\theta_h$. It is convenient to define $R$ as the linear $(J+1)\times(J+1)$ transformation matrix such that $\mathbf{\theta}=R\mathbf{S}$ and to define the $N\times(J+1)$ matrix $C$ that consists of the $C_{ij}$ terms defined above. Finally, we define the matrix $D$ as $D=CR$. Then the log-likelihood can then be expressed as:

\begin{eqnarray}\label{loglikelihoodallN}
 l(S,\beta) = \sum_{i=1}^N  l_i(S,\beta) 
&=&  \sum_{i=1}^N \Bigg[ (1-M_i)\Delta_i \log \Bigg( \sum_{j=1}^{V_i} D_{ij} (S_j)^{\exp(x_i'\beta)} \Bigg) + \nonumber \\ &&  (1-M_i)(1-\Delta_i) \log \Bigg( \sum_{j=V_i+1}^{J+1} D_{ij} (S_j)^{\exp(x_i'\beta)}  \Bigg) 
+ \nonumber \\ && M_i \log \Bigg( \sum_{j=1}^{J+1} D_{ij} (S_j)^{\exp(x_i'\beta)}  \Bigg)\Bigg].
\end{eqnarray}

We can solve for the unknown vector of parameters $\psi$ using standard maximum likelihood estimation. Define the score function $U_i({\psi})=\frac{\partial l_i(S,\beta) }{\partial \psi}$, where $\mathbf{S} = (S_1,S_2,...,S_{J+1})'$ and $\mathbf{\psi}$ is the $(p+J+1) \times 1$ parameter vector $\mathbf{[\beta,S]}$. Let $\hat{\psi}$ denote the solution to the equations $\sum_{i=1}^N U_i({\psi})=0$. The covariance matrix can be found by inverting the Hessian matrix. 
\subsection{Survey Design and Probability
Sampling Weights}\label{surveymethods} 
In this section, we extend our proposed method that uses both auxiliary and gold standard outcomes to accommodate data from a complex survey sampling design, such as HCHS/SOL, that may includes cluster-based probability sampling. We develop a weighted analogue of our log-likelihood function from equation \ref{loglikelihoodallN}. Later, we outline how one might use a sandwich variance estimator to address within-cluster correlation and stratification.

Define $\pi_i$ as the probability that subject $i$ will be included in a sample, which we assume is known from the survey design. Subjects are sampled with probability $\pi_i$ from a population of size $N_{POP}$, resulting in a sample of size $N$.  Design-based inference makes the assumption that a subject sampled with a probability $\pi_i$ represents $1/\pi_i$ subjects in the total population. \citep{lumley2011complex} Thus, $1/\pi_i$ becomes the sampling weight reflecting unequal probability of selection into the sample, which will be included in the weighted log-likelihood and score functions. The weighted log-likelihood equation becomes $l_{\pi}(S,\beta) = \sum_{i=1}^N \frac{1}{\pi_i}  l_i(S,\beta) = \sum_{i=1}^N  \widecheck{l}_i(S,\beta).$ We can then use standard maximum likelihood theory to solve the corresponding weighted estimating equation $\sum_{i=1}^N \widecheck{U}_{i}({\psi})=\sum_{i=1}^N \frac{1}{\pi_i} U_{i}({\psi})=0$ for our vector of unknown parameters, $\psi$. To compute the variance for our estimator that addresses within-cluster correlation and stratification, we consider the implicit differentiation method proposed by Binder (1983).\citep{binder1983variances} Using a Taylor series linearization, the sandwich estimator for the asymptotic variance of $\hat{\psi}$ can be calculated as $\hat{\text{var}}[\hat{\psi}] \approx \left(\sum_{i=1}^N \frac{\partial \widecheck{U}_{i} (\hat{\psi})}{\partial \psi}\right)^{-1} \hat{\text{cov}}\left[\sum_{i=1}^N \widecheck{U}_{i} (\hat{\psi}) \right] \left(\sum_{i=1}^N \frac{\partial \widecheck{U}_{i} (\hat{\psi})}{\partial \psi}\right)^{-1}$.
 Regularity conditions required for the consistency of $\hat{\text{var}}[\hat{\psi}]$ are stated in Binder (1983).\citep{binder1983variances} This variance estimate can easily be computed in R by applying vcov() to the svytotal() function from the survey package and providing the estimator's influence function as well as the survey design. \citep{lumley2011complex}
\subsection{Regression Calibration to Adjust for Covariate Measurement Error}\label{regcalsection} 
 Regression calibration is a popular analysis method for correcting bias in regression parameters when exposure variables are prone to error. \citep{prentice1982covariate,shaw18} We will now outline how to use regression calibration with our proposed estimator in the setting of a complex sampling design. 

Assume $(X,Z)$ is a $(p+q)$-dimensional covariate in the outcome model of interest, where $X_i$ is a $p$-dimensional vector that cannot be observed without error and $Z_i$ is a $q$-dimensional vector of observed, error-free covariates. Assume instead of $X_i$, we observe $X_i^*$, the corresponding error-prone  $p$-dimensional vector.  To implement regression calibration, we build a calibration model for $\hat{X}=E(X|X^*,Z)$ and substitute this predicted value for the unknown, unobserved true exposure $X$ in our outcome model. \citep{prentice1982covariate,keogh2020stratos}

\subsubsection{Measurement Error Model}\label{MEmodelsection} 

We assume that the error-prone $X_i^*$, is linearly related with the target exposure $X_i$ and other error-free covariates $Z_i$:

\begin{equation}\label{regcalmodel}
    X_i=\delta_{(0)}+\delta_{(1)}X^{*}_i+\delta_{(2)}Z_i+\zeta_i,
\end{equation}

\noindent where $\zeta_i$ is a random error term that has mean zero and variance $\sigma_{\zeta_i}^2$ and is independent of $X^*_i$ and $Z_i$. Equation (\ref{regcalmodel}) is referred to as the \textit{calibration model}. For ease of presentation, we assume $p=1$. It follows that the observed, error-prone exposure $X_i^{*}$ conforms to the linear measurement error model: $X_i^*=\alpha_{(0)}+\alpha_{(1)}X_i+\alpha_{(2)}Z_i+e_i$, where the random error $e_i$ is independent of $X_i$ and $Z_i$ and has mean zero and variance $\sigma_{e_i}^2$. \citep{keogh2020stratos} This error model has been commonly applied to model the error in the self-reported dietary intake exposures observed in our motivating example from the HCHS/SOL. \citep{keogh14}  Regression parameters in our calibration model are identifiable if, in a subset, we observe either the true exposure, $X_i$, or a second error-prone observation $X_i^{**}$ with classical measurement error,  i.e., where $X_i^{**}=X_i+\epsilon_i$, where $\epsilon_i$ is random error that is independent of all variables, with mean 0 and variance $\sigma_{\epsilon_i}^2$. In many settings, it is more common to observe  $X_i^{**}$ in the ancillary data, which we call a calibration subset.  We will assume a subset is available in which we observe $X_i^{**}$. Note that observing the true exposure $X_i$ is a variation of observing $X_i^{**}$ in which the measurement error variance $\sigma_{\epsilon_i}^2$ is equal to 0, and such a subset is referred to as a validation subset. In some applied settings, the error-prone measure $X_i^*$ in the main data may only have classical measurement error, a special scenario where  $\alpha_{(0)}=\alpha_{(2)}=0$ and $\alpha_{(1)}=1$ in the linear measurement error model. In this case, a replicate measure in the ancillary data (typically called a reliability subset) will ensure that the parameters in the calibration model are identifiable. 

With the assumed calibration subset, we can regress $X_i^{**}$ on the error-prone exposure, $X_i^*$, and other covariates of interest $Z_i$ to fit the model $X^{**}_i=\delta_{(0)}+\delta_{(1)}X_i^{*}+\delta_{(2)}Z_i+W_i$,
\noindent  where $W_i$ is random, mean 0 error with variance $\sigma_{W_i}^2=\sigma_{\zeta_i}^2+\sigma_{\epsilon_i}^2$. The error term $W_i$ in this model now incorporates extra variability introduced by the error in $X_i^{**}$.
\subsubsection{Applying Regression Calibration to the Outcome Model}\label{applyingRegCal} 
Assuming that the measurement error models described above hold, we can use the predicted values from our calibration model to substitute the first moment $\hat{X}_i=E(X_i|X_i^*,Z_i)$ in place of $X_i$ in our outcome model.  Regression calibration is exact in linear models; however, this approach is only an approximate method with some bias in non-linear outcome models. \citep{carroll2006measurement} Regression calibration has been observed to perform well in various settings, including when the regression parameter corresponding to the error-prone covariate is of modest size and when the event under study is rare. \citep{prentice1982covariate,buono10} Additionally, regression calibration has been been shown to work well under these same settings when also correcting for errors in time-to-event outcomes. \citep{boe2021approximate}

As we described in section \ref{surveymethods}, variance estimation for data from a complex survey design often requires extra steps to address within-cluster correlation. When regression calibration is applied, variance estimates from the outcome model need to be adjusted further to account for the extra uncertainty added by the calibration model step. We adopt the variance estimation approach proposed by Baldoni et al. (2021)\citep{baldoni2021use}, in which the expected value of the latent true exposure is multiply imputed for all individuals by repeatedly sampling the calibration model coefficients required to estimate $\hat{X}_i$. New calibration coefficients can be sampled using either (1) their estimated asymptotic parametric distribution or (2) bootstrap resampling. At each step of the imputation, the outcome model is re-fit using the newly calibrated values. Using this approach, the final estimate of the variance of the $j$th regression coefficient $\hat{\beta}_j$ can be computed as $\hat{V}_j^*=\frac{1}{M}\sum_{m=1}^M \hat{V}_j^{(m)}+\frac{1}{M-1}\sum_{m=1}^M\Big(\hat{\beta}_j^{(m)}-\bar{\hat{\beta}}_j\Big)^2$, where $\bar{\hat{\beta}}_j=\frac{1}{M}\sum_{m=1}^M\hat{\beta}_j^{(m)}$ and $\hat{\beta}_j^{(m)}$ and $\hat{V}_j^{(m)}$ represent the estimated $j$th regression coefficient and its estimated variance, respectively, using the $m$-th completed data set with $m=1,\ldots,M$. Further details on variance estimation can be found in Baldoni et al. (2021)\citep{baldoni2021use} and code for implementing them is available on GitHub at \url{https://github.com/plbaldoni/HCHSsim}. 
\subsection{Asymptotic Theory}\label{asymptotictheoryintro} 
We assume the regularity conditions of Foutz (1977) \citep{foutz1977unique} and apply the techniques of Boos and Stefanski (2013) \citep{boos2013essential} for verifying asymptotic normality of standard maximum likelihood estimators to establish the asymptotic properties of the proposed estimator.  In section S3 of the Supplementary Materials, we outline regularity conditions for the following three settings: (1) the proposed method estimator is applied to data from a simple random sample from the population; (2) the proposed method estimator is extended to accommodate data from a complex survey design; and (3) the proposed method estimator is extended to incorporate regression calibration in the presence of complex survey data.
\section{Numerical Study}\label{section3}
We now present a simulation study conducted to assess the numerical performance of the proposed method compared to the standard  discrete proportional hazards model approach for the gold standard time-to-event outcome. Regression coefficients for the standard approach are obtained by fitting a generalized linear model with a binary response and complementary log-log link. \citep{hashimoto2011regression} We explore various settings to show when the proposed estimator improves over the standard interval-censored approach in terms of statistical efficiency. In particular, we vary the probability that the gold standard indicator $\Delta_i$ is missing for some subjects, the censoring rate ($CR$) of the latent true event time at the end of study (i.e. if $\Delta_i$ had been observed for all subjects), and the sample size, N. Additionally, we vary the missingness rate of our auxiliary outcome variable and consider different values for our true regression parameter of interest, $\beta$, different distributions of our simulated event times, and different values of sensitivity and specificity of the auxiliary data.
\subsection{Simulation Setup}
We first consider a set of simulations assuming a simple random sample. We simulate a single covariate of interest from either a gamma distribution with shape and scale parameters of 0.2 and 1, respectively (denoted Gamma$(0.2,1)$) or a normal distribution with mean and variance parameters 0.2 and 1 (denoted Normal$(0.2,1)$). We assume the proportional hazards model. We fix the true log hazard ratio at $\beta=\log(1.5)$ to represent a regression coefficient of moderate size. Later, we set $\beta=\log(3)$ to see how increasing the magnitude of our regression coefficient changes our efficiency gains. Additionally, we conduct simulations with $\beta=0$ to check type I error rates, where $\alpha=0.05$. All simulations were run in R version 4.1.0. \citep{citeR}

True event times were generated from a continuous time exponential distribution. We simulated a follow-up schedule with four fixed visit times at which we collect the auxiliary outcome variables. We assume that at year four, a gold standard outcome variable is also recorded. To obtain average censoring rates ($CR$) for the latent true event of 0.9, 0.7, and 0.5, we considered baseline $\lambda_b$ parameters of 0.023, 0.08, and 0.17, respectively, and simulated our event times using parameter $\lambda=\lambda_b\exp(x_i'\beta)$. We discretize the continuous event times by binary event indicators for each visit time, then use sensitivity and specificity values to ``corrupt" this variable, resulting in the vector of error-prone auxiliary outcomes, $\mathbf{Y_i^*}$. We varied the accuracy of our auxiliary data by considering scenarios where sensitivity $=0.90$ and specificity $=0.80$, as well as sensitivity $=0.80$ and specificity $=0.90$.  
To simulate scenarios in which the gold standard outcome $\Delta_i$ is not observed for some subjects ($M_i=1$), we vary the missingness rate ($MR$) of $\Delta_i$ at 0, 0.2 and 0.4. To simulate this missingness, we generated $N$ variables $U_i$ from a Uniform(0,1) distribution and then let $\Delta_i$ be missing for each subject if $U_i<MR$. We vary the sample size between $N=1000$ and $N=10,000$ subjects. When $MR=0.0$, these sample sizes are exact for the proposed approach and the no auxiliary data approach. When $MR>0.0$, $N=1000$ and $N=10,000$ represent the sample sizes for the proposed approach, but the true sample sizes for the standard (no auxiliary data) approach are smaller due to missingness in the gold standard indicator $\Delta$.  For all settings, we conducted 1000 simulation iterations. 

We then performed a set of simulations with similar settings, except we sought to examine the performance of the proposed method with data having the structure of a complex survey design.  Code for this set of simulations was developed and described by Baldoni et al. (2021)\citep{baldoni2021use} and is available on GitHub at \url{https://github.com/plbaldoni/HCHSsim}. Briefly, this simulation pipeline creates a superpopulation of nearly 200,000 individuals in 89,777 households, across 376 block groups, and 4 geographic strata and then for each simulation iteration drew survey samples from it using a stratified three-stage sampling scheme. The resulting simulated data sets include sampling weights, stratification variables, and cluster indicators. To simulate our gamma covariate for this set of simulations, we considered different shape and scale parameters for the four strata: shape$_1$ = 0.25, scale$_1$  = 1.25;  shape$_2$  = 0.15, scale$_2$ = 0.75;  shape$_3$ = 0.30, scale$_3$  = 1.50;  shape$_4$ = 0.10, scale$_4$  = 0.50. For each block group $g$ within a certain stratum $s$, we created additional covariate differences by simulating variables $\omega_{gs}$ from a Uniform($-0.15*\textrm{shape}_s,0.15*\textrm{shape}_s$) and $\rho_{gs}$ Uniform($-0.15*\textrm{scale}_s,0.15*\textrm{scale}_s$)  distribution for $s=1\ldots,4$. Then, the covariate for an individual in block group $g$ and stratum $s$ was simulated from a Gamma$(\textrm{shape}_s+\omega_{gs},\textrm{scale}_s+\rho_{gs})$ distribution. To illustrate the performance of our method under the complex survey design with a normally distributed covariate, we also considered variables $X_i \sim$ Normal$(\textrm{shape}_s+\omega_{gs},\textrm{scale}_s+\rho_{gs})$. All other settings, including setting $\beta=\log(1.5)$ and the generation of the event times and the missingness in the gold standard, were kept the same between the random sample and complex survey for this set of simulations. Due to the randomness introduced by the complex survey sampling setting, we cannot fix the total number of individuals selected for a simulated sample, but we aimed for sample sizes of approximately $N=1000$ and $N=10,000$ as in prior tables. 

We conducted one additional simulation that aimed to mimic the HCHS/SOL study, which included error-prone covariates. We aimed for an average sample of approximately $12,987$ in order to approximate the number of HCHS/SOL cohort subjects without baseline diabetes. We assumed eight fixed visit times at which the auxiliary outcome was recorded, with a simulated gold standard occurring at year four. Missingness in the gold standard indicator at year four was set at $MR=0.29$, the censoring rate was fixed at roughly $CR=90\%$, and the auxiliary data missingness rate was approximately $0.20$. We simulated 3 covariates of interest: $X$, $Z_1$, and $Z_2$ to represent dietary intake, age, and body mass index (BMI), respectively. These covariates were simulated following the data generation structure of Baldoni et al. (2021)\citep{baldoni2021use}, where each subject's sex (male, female) and Hispanic/Latino background (Dominican, Puerto Rican, and other) were first simulated from a multinomial distribution. Next, self-reported dietary intake, age, and BMI were simulated for each combination of sex and Hispanic background following a multivariate normal distribution, with means and covariance matrices estimated from the HCHS/SOL Bronx field center data. We set $\beta_1=\log(1.5)$, $\beta_2=\log(0.7)$, $\beta_3=\log(1.3)$. To simulate an error-prone covariate $X^*$, we use the linear measurement error model, $X^*=\alpha_{(0)}+\alpha_{(1)}X+\alpha_{(2)}Z_1+\alpha_{(3)}Z_2+e$, where $\alpha_{(0)}=0.05$, $\alpha_{(1)}=0.50$, $\alpha_{(2)}=0.003$, and $\alpha_{(3)}=0.0009$. We assumed $e \sim N(0,\sigma_{e}^2)$  and used a $\sigma_{e}^2$ value of 0.389. To represent the biomarker subset, we take a random sample of $450$ participants on which we observe a measure with classical error, simulated as $X^{**}=X+\epsilon$, where $\epsilon \sim N(0, \sigma_{\epsilon}^2)$ and $\sigma_{\epsilon}^2=0.019$. These values of  $\alpha_{(0)}, \alpha_{(1)},\alpha_{(2)}, \alpha_{(3)}$, $\sigma_{e}^2$  and $\sigma_{\epsilon}^2$ were chosen based on parameters fit for the self-reported and recovery biomarker measurements for protein density in the HCHS/SOL data. \citep{mossavar2015applying}

For all simulation settings we conducted 1000 simulation iterations and report median percent ($\%$) biases, median standard errors (ASE), empirical median absolute deviation (MAD), 95\% coverage probabilities (CP), and median relative efficiencies (RE), calculated as the median of the ratio of the estimated variance of the proposed method estimator to the estimated variance of the standard approach estimator. R code used to run our simulations can be found on GitHub at \url{https://github.com/lboe23/AugmentedLikelihood}. 
\subsection{Simulation Results}
In Tables \ref{table1sims}-\ref{t1errortable}, we present results for the proposed method compared to the standard interval-censored approach without auxiliary data. Table \ref{table1sims} shows results for the simple random sample with the  regression parameter of interest $\beta=\log(1.5)$ and a gamma distributed covariate. The proposed method performs well, maintaining an absolute median percent bias of under 2\% for all settings and achieving nominal coverage for a 95\% confidence interval. We also see that our variance estimator is working properly, as our ASE values closely approximate the MAD values. We note that substantial efficiency gains (1.2-69.9\%) result from incorporating auxiliary data into the analysis. Our method shows larger efficiency gains when the missingness rate, $MR$, for the gold-standard indicator $\Delta$ is higher and when the censoring rate of the latent true event time at the end of study $CR$ is lower. Table S1 in the Supplementary Materials shows a benchmark for comparing the relative efficiency gains from the proposed method to the relative efficiency gains achieved if the gold standard were available at all four visit times. We can directly compare the relative efficiency improvements from the final column of Table S1 to those in the final column of Table \ref{table1sims} to see that for these particular settings, our method retains nearly 90\% of the the ideal relative efficiency. 

In Table S2 from the Supplementary Materials, we change the sensitivity and specificity values for the auxiliary outcome and let $Se=0.90$ while $Sp=0.80$. We see that our method still performs well with these alternate values for $Se$ and $Sp$ in terms of mean percent bias, standard error estimation, and coverage probability. When $MR=0.0$, relative efficiencies are similar between Table \ref{table1sims} ($Se=0.80$, $Sp=0.90$) and Table S2 ($Se=0.90$, $Sp=0.80$). For example, when $CR=0.50$ and $N=10,000$, we have an efficiency gain of $1.186$ in Table \ref{table1sims} and an efficiency gain of $1.178$ in table Table S2. However, when $MR>0$, we notice more substantial efficiency gains for Table \ref{table1sims}, where sensitivity is lower and specificity is higher, e.g. 1.677 vs. 1.549 for $MR=0.4$, $CR=0.50$ and $N=10,000$. 

Table \ref{table3sims} shows the results when the  covariate of interest follows a normal distribution. Relative efficiencies in this table range from 0.1\% to 39.8\%, indicating that efficiency gains are not as high for a normally distributed covariate. We also assess the gains in relative efficiency for the proposed method over the standard interval-censored approach for $\beta=\log(3)$ in Table S3 in the Supplementary Materials. Increasing the magnitude of our regression coefficient leads to much larger increases in relative efficiency, ranging from 15.5\% to 117\%.

Table \ref{weightedtablegammaMED} presents results for data simulated from a complex survey. In all scenarios, the weighted proposed estimator has minimal finite sample bias. The sandwich variance estimator performs unfavorably in some settings for both the proposed and standard method, with coverage as low as 89.9\%, particularly when the sample size is small ($N=1000$) or the $CR$ is high. We note, problematic finite sample performance of the sandwich variance has been observed in other settings where the number of observed events is modest and/or the covariate is from a skewed distribution \citep{carroll1998sandwich}. For all settings, relative efficiency gains are observed to be quite high for the proposed method, ranging from 0.9\% to 60.9\%. In Table S4 from the Supplementary Materials, we show results for data simulated from a complex survey design using a normally distributed covariate. With a symmetrical covariate, the sandwich variance estimator performs better, achieving empirical MADs that more closely resemble the ASEs and obtaining coverage closer to the nominal 95\% level. However, as we observed for the random sample case, relative efficiency gains are not as large (1\%-45.5\%) using a normally distributed covariate.

We present results for the simulation that mimic the data  structure and complex survey design of the HCHS/SOL study in Table \ref{HCHSSIMS}. Median percent bias is -$0.859\%$ for the proposed estimator and we see that applying the multiple imputation-based variance correction approach leads to well-behaved sandwich standard errors. We estimate a relative efficiency gain of 44.2\%, suggesting that our approach can lead to substantial variance reductions under the data structure and measurement error settings similar to that of the HCHS/SOL cohort. Finally, we assess type I error results in Table \ref{t1errortable}. Type I error rates ranged from 0.033 to 0.065 for different values of $MR$, $CR$, and $N$, indicating that type I error is preserved in the proposed method for all observed settings.

\section{Hispanic Community Health Study/Study of Latinos (HCHS/SOL) Data Example}\label{section4}

\subsection{HCHS/SOL Study Description}

The Hispanic Community Health Study/Study of Latinos is an ongoing multicenter community-based cohort study of 16,415 self-identified Hispanics/Latino adults aged 18-74 years recruited from randomly selected households at 4 locations in the United States (Chicago, Illinois; Miami, Florida; Bronx, New York; San Diego, California). Households were selected using a stratified 2-stage area probability sample design. The sampling methods, design, and cohort selection for HCHS/SOL have been described previously. \citep{sorlie2010design, lavange2010sample} The study was designed to identify risk factors for chronic diseases including diabetes and to quantify morbidity and all-cause mortality.  Prevalent diabetes was recorded using a biomarker-defined reference standard at the baseline, in-person clinical examination visit (2008-2011). The study design was such that all participants were scheduled to be assessed for incident diabetes using (1) a biomarker-defined reference standard at a second clinic visit (visit 2) 4-10 years after baseline, and (2) annual telephone follow-up assessments recorded by self-report. Participants have up to eight annual telephone follow-up calls. We found that most ($>97\%$) participants' follow-up call dates rounded to exactly one year from the date of their prior call, so we used the assigned annual follow-up times to define the boundaries of the follow-up intervals. Follow-up time was divided into 9 possible intervals.  To define the observation time for the reference standard at visit 2, we rounded the time between baseline and the second clinic visit to the nearest year. Visits that occurred after year 8 (1.51\% of all visits) were rounded down in order to preserve the visit schedule with 9 intervals. For the interval-censored, no auxiliary data approach, we assumed that visit 2 occurred at the same time for all participants that had the reference standard available. Note we made this simplifying assumption due to the lack of available software to handle the complex survey design for the interval-censored proportional hazards model. We used this as a comparative analysis that did not use auxiliary data.

We applied the proposed method to assess the association between energy, protein and protein density (percentage of energy from protein) dietary intakes and the risk of diabetes in HCHS/SOL using both the self-reported diabetes outcome (auxiliary data) and the reference standard. The dietary exposure variables were recorded using an error-prone, self-reported 24-hour recall instrument that is believed to follow to the linear measurement error model. A subset of 485 HCHS/SOL participants were enrolled in the Study of Latinos: Nutrition and Physical Activity Assessment Study (SOLNAS). \citep{mossavar2015applying} The SOLNAS subset included the collection of objective recovery biomarkers that conform to the classical measurement error model and therefore can be used to develop calibration equations for the self-reported dietary intake variables. 

This work was motivated by more detailed, ongoing research looking to understand the relationship between several dietary factors and risk of chronic diseases, including diabetes and cardiovascular disease, in the HCHS/SOL cohort. The proposed method is applied to a random subset of $8,200$ eligible participants, which is half of the original HCHS/SOL cohort ($N=16,415$). Eligibility included being diabetes-free at baseline and having complete covariate data. Details on eligibility and the selection of our random subset are provided in section S4 of the Supplementary Materials. Our calibration models for dietary energy, protein, and protein density  included age, body mass index (BMI), sex, Hispanic/Latino background, language preference, income, and smoking status. We fit the calibration equation by regressing the biomarker value $(X^{**})$ on the corresponding self-reported measure and other covariates. We compared self-reported diabetes and the reference standard at baseline to determine that self-reported diabetes in HCHS/SOL has a sensitivity of 0.61 and a specificity of 0.98. We also conduct a sensitivity analysis in which we use a sensitivity of 0.77 and a specificity of 0.92, which are the measures of agreement computed using self-reported diabetes and the reference standard diabetes measure at visit 2.

All analyses accounted for the HCHS/SOL complex survey design. To fit the model for the interval-censored reference standard diabetes measure from visit 2, we used the svyglm() function from the survey package in R. \citep{lumley2011complex} To apply our proposed approach, we maximized the weighted log-likelihood that included HCHS/SOL sampling weights, and obtained design-based standard errors using the approach outlined in section \ref{surveymethods}. 
The models for both approaches are fit $M$ times, once for each of the newly predicted intake values $\hat{X}_i^{(m)}$ from multiple imputation. The final variance estimate is computed using the approach described in section \ref{applyingRegCal}. We chose $M=25$ imputations for our analysis. In both models, we used biomarker calibrated values of dietary energy, protein, and protein density on the log scale. Both risk models were also adjusted by the standard risk factors included in the calibration equations. We present hazard ratios (HR) and 95\% confidence intervals (CI) associated with a 20\% increase in consumption.
\subsection{Results}
Of the 8,200 randomly selected participants, 5,922 (72.2\%) had the reference standard diabetes status variable available at visit 2. Of participants who had visit 2 data, 5 (0.1\%) participants returned to the clinic four years post-baseline, 1490 (25.2\%) returned after five years, 3294 (55.6\%) returned after six years, 739 (12.5\%) returned after 7 years, and 394 (6.7\%) returned after 8 years. Using the reference standard, 623 (10.5\%) of the participants with visit 2 data had incident diabetes. 

Table \ref{tableHR} shows results from applying the proposed method and the standard, no auxiliary data method to the HCHS/SOL data. The HR (95\% CI) for a 20\% increase in energy intake was 1.20 (0.47, 3.11) for the proposed approach compared to 1.20 (0.41, 3.82) for the no auxiliary data method. For energy, we observe a relative efficiency gain of 27\% by using the proposed method. While the estimated standard error for the no auxiliary data approach is larger compared to that of the proposed method, incident diabetes is not significantly associated with energy intake in either approach. For protein, the HR (95\% CI) for a 20\% increase in intake using the proposed method is estimated to be 1.30 (0.82, 2.06). Comparatively, we estimate an HR (95\% CI) of 1.37 (0.74, 2.51)  using the no auxiliary data approach, and estimate a corresponding relative efficiency gain of 74\% using the proposed method. When the proposed method is applied, the HR for a 20\% increase in protein density is estimated to be 1.01 (1.00, 1.02), compared to a HR of 1.01 (1.00, 1.03) for the no auxiliary data method. Our estimated relative efficiency gain using the proposed method over the standard approach is 63\% when looking at protein density. We note that this large efficiency gain was from relatively small absolute changes on the log-hazard scale.

In Table S5 of the Supplementary Materials, we present results from a sensitivity analysis that applies the proposed method using sensitivity and specificity values estimated at visit 2 ($Se=0.77, Sp=0.92)$. For this investigation, we use the same subset of $8,200$ HCHS/SOL participants as in the primary analysis. We observe that changing the sensitivity and specificity values does not qualitatively change our results for any of the dietary intakes under study. 

\section{Discussion}\label{discussion}
In large cohort studies like HCHS/SOL, gold or reference standard outcome variables may be less readily available than error-prone auxiliary outcomes. We have introduced a method that leverages all available data by incorporating error-prone auxiliary variables into the analysis of an interval-censored outcome. We developed methods for both a simple random sample and complex survey design for the case of time-independent covariates. Our results suggest that making use of auxiliary outcome data may often lead to a considerable improvement in the efficiency of parameter estimates, particularly when the gold standard outcome is missing for a subset of study participants. We illustrate the practical use of our approach in a complex survey design by applying the proposed method to the HCHS/SOL study to assess the association between energy, protein, and protein density intake and the risk of incident diabetes, while adjusting for error in the self-reported exposure. In HCHS/SOL, the reference standard diabetes outcome variable was not practical to obtain annually, while self-reported diabetes status was easily attainable. This data example served as a compelling setting for which our method could contribute, reducing the estimated variance by up to 74\%. In settings with substantial measurement error, where variance estimates can be quite large, relative efficiency improvements are extremely important and may inform cost reductions for future studies.

In the HCHS/SOL study, we observe a special case of interval-censored data in which the reference standard outcome is only observed at one time point. This type of data is often called current status data, or case I interval-censored data. \citep{zhang2010interval} In our data example, the current status data arise due to the study design, as the reference standard outcome  was scheduled to be recorded only once at a predetermined time point post-baseline. However, under the discrete time framework in which there is a common set of assessment times for all individuals, our method could be easily adapted to accommodate a reference standard status variable recorded at multiple time points. For the continuous time setting, future work is needed to consider how our estimation methods for interval-censored data could be extended. Several approaches have been applied for the analysis of continuous time interval-censored data, many of which have been shown to be computationally complex. \citep{zeng2016maximum, zhang2010spline, lindsey1998methods} These methods, however, have not yet been adapted to handle error-prone and validated outcomes. A further extension would be to consider approaches able to handle time-varying covariates.

The application of the proposed method required defining a set of common, discrete visit times across participants to avoid the curse of dimensionality. We used the assigned annual visit times to define the boundaries of the visit intervals, thus ignoring that the annual visit may not occur on the participant's exact anniversary date. We deemed this appropriate because the observed visit times were generally quite close to the anniversary times. In other settings,  where the fluctuations in visit times are more extreme, one might consider dividing time into smaller intervals. For this approach, the choice of intervals will require us to consider to what the extent the data can support estimating the increased number of nuisance parameters from a finer grid. With discrete data, we must often make a pragmatic compromise that  balances the bias induced from rounding event times and the problems that may arise from a large number of parameters. Extending our methods in a way that does not restrict the number of possible visit times and allows for more parameters to be stably estimated need further investigation.

One potential limitation of our analysis of the HCHS/SOL data was the assumption of constant sensitivity and specificity across visit times, as there was some apparent disagreement between these measures of accuracy at baseline compared to visit 2. We hypothesize that this difference in agreement is primarily a result of a larger lag time since the previous gold standard test at baseline compared to follow-up visits, but could also result from missing data in the reference measure at visit 2 that may impact the sensitivity and specificity values. We conducted a sensitivity analysis to explore how using visit 2 rather than baseline values of sensitivity and specificity may impact the results of our HCHS/SOL data analysis. In this example, incorporating slightly different measures of accuracy of the self-reported auxiliary outcome data did not substantially impact our results. However, we note that this may not always be the case, especially for more extreme changes in sensitivity and specificity. For many real data settings, it may be unreasonable to assume that the sensitivity and specificity of error-prone outcomes are time-invariant. Future methods might explore the possibility of incorporating time-varying values of sensitivity and specificity. A second potential limitation was our assumption that the gold standard outcome was missing completely at random. Using our proposed method for the complex survey design, we anticipate an extension could be readily developed to handle the missing at random case with the use of inverse probability weighting.

In our numerical study, we noticed that the sandwich variance estimator had some coverage issues in smaller sample settings using both the proposed method and the standard no auxiliary data approach. While the sandwich variance estimator performed better with a normally distributed covariate, we noticed some numerical challenges when the covariate of interest had a long-tailed distribution (e.g. the gamma distribution). The numerical limitations of the sandwich variance estimator for complex survey data in non-linear models have been discussed previously. Bias in the sandwich estimator may be encountered with smaller sample sizes and rare outcomes, particularly for a covariate with a heavy-tailed distribution, since in these settings, the variability of regression parameters is underestimated. \citep{carroll1998sandwich, rogers2015modification} Despite these limitations, the sandwich estimator may be reasonable, as coverage remained above 89\%, got closer to 95\%  in large samples, and it is very practical to implement.

There are several methods for variance estimation that may be considered when applying regression calibration. Further steps are typically required to incorporate the extra uncertainty added by the calibration model and make valid inference on the parameter of interest. In practice, the bootstrap estimator is often used due to its simple implementation.  For cases in which the data are from a complex survey design like HCHS/SOL, additional considerations are needed to account for aspects of the sampling design and the standard bootstrap variance estimator may be less straightforward to apply. While the multiple imputation approach of Baldoni et al. (2021)\citep{baldoni2021use} can be applied in these scenarios, we observed some instances of over-coverage of the 95\% confidence intervals using these variance estimators (data not shown). This issue is discussed by Baldoni et al. (2021)\citep{baldoni2021use} and is believed to be attributed to instability introduced by multicollinearity in the simulated data. Future work may consider alternative variance estimation strategies in the presence of regression calibration and the complex survey setting, such as a sandwich variance estimator obtained by stacking the calibration and outcome model estimating equations.
\section*{Acknowledgements}
The authors would like to thank the investigators of the HCHS/SOL study for the use of their data. A list of HCHS/SOL investigators, managers and coordinators by field center can be found here: \url{https://sites.cscc.unc.edu/hchs/Acknowledgement}.

\section*{Data Availability Statement}
The data used in this paper was obtained through submission and approval of a manuscript proposal to the Hispanic Community Health Study/Study of Latinos Publications Committee, as described on the HCHS/SOL website. \citep{SOLdatacite} For more details, see \url{https://sites.cscc.unc.edu/hchs/publications-pub}. 

\section*{Declaration of conflicting interests}
The authors declared no potential conflicts of interest with respect to the research, authorship and/or publication of this article.

\section*{Funding}
This work was supported in part by NIH grant R01-AI131771. 

\section*{Supporting Information}
The Supplementary Materials corresponding to this paper are available online. R code for our simulations and a sample data analysis that applies the proposed method to a simulated data set with similar features to the HCHS/SOL is available on GitHub at

\noindent \url{https://github.com/lboe23/AugmentedLikelihood}. Additionally, Section S1 of the Supplementary Materials provides R code illustrating how to apply  (1) the proposed method and (2) the standard, no auxiliary data method to a simulated data set.

\newpage 

\begin{table}
\centering
      \begin{threeparttable}[t]
            \caption{Simulation results are shown for exponential failure times assuming the Cox proportional hazards model with $X\sim Gamma(0.2,1)$ and $\beta=\log(1.5)$. The median percent (\%) bias, median standard errors (ASE), empirical median absolute deviation (MAD) and coverage probabilities (CP) are given for 1000 simulated data sets for the proposed method and the standard interval-censored approach that does not incorporate auxiliary data. Here, $Se = 0.80$ and $Sp = 0.90$ for the auxiliary data.}\label{table1sims}
   
      \begin{tabular}{cccccccccccccc}
\hline
\multicolumn{3}{l}{} & \multicolumn{4}{c}{Proposed}  & \multicolumn{4}{c}{No Auxiliary Data} & \multicolumn{1}{c}{ } \\ \cmidrule(r){4-7} \cmidrule{8-11}  $MR$\tnote{1}  & $CR$\tnote{2} & $N$\tnote{3}   & \% Bias & ASE & MAD & CP & \% Bias & ASE & MAD & CP & \multicolumn{1}{c}{RE\tnote{4}}  \\ 

\hline
 0.0 & 0.9 & 1000 &   -0.958 & 0.159 & 0.150 & 0.956 & -1.402 & 0.160 & 0.155 & 0.951 & 1.012 \\

 & & 10,000 &    1.351 & 0.048 & 0.050 & 0.947 & 1.279 & 0.048 & 0.051 & 0.951 & 1.010 \\

& 0.7 & 1000 &   0.824 & 0.103 & 0.100 & 0.947 & 0.614 & 0.107 & 0.106 & 0.950 & 1.053 \\

 & & 10,000 &    0.543 & 0.032 & 0.032 & 0.944 & 0.398 & 0.033 & 0.034 & 0.947 & 1.070 \\

 & 0.5 & 1000 &    1.923 & 0.091 & 0.088 & 0.943 & 2.020 & 0.099 & 0.102 & 0.947 & 1.182 \\

 & & 10,000 &   0.521 & 0.028 & 0.029 & 0.946 & 0.382 & 0.031 & 0.034 & 0.951 & 1.186 \\

 0.2 & 0.9 & 1000 &    -1.071 & 0.172 & 0.170 & 0.957 & -0.378 & 0.181 & 0.183 & 0.951 & 1.072 \\

 & & 10,000 &  1.199 & 0.052 & 0.050 & 0.958 & 0.769 & 0.054 & 0.055 & 0.952 & 1.087 \\

 &  0.7 & 1000 &    1.333 & 0.109 & 0.106 & 0.953 & 0.377 & 0.120 & 0.116 & 0.954 & 1.184 \\ 
  
 & &  10,000 &    0.713 & 0.034 & 0.035 & 0.942 & 0.332 & 0.037 & 0.038 & 0.946 & 1.206 \\

 &  0.5 & 1000 &    1.798 & 0.095 & 0.095 & 0.945 & 2.084 & 0.111 & 0.116 & 0.947 & 1.363 \\

 &   & 10,000 &    0.534 & 0.029 & 0.030 & 0.945 & 0.247 & 0.034 & 0.036 & 0.952 & 1.370 \\ 
  
 0.4 & 0.9 & 1000 &    0.256 & 0.189 & 0.188 & 0.959 & 1.178 & 0.213 & 0.222 & 0.959 & 1.195 \\

 & & 10,000 &     1.444 & 0.056 & 0.057 & 0.951 & 2.122 & 0.062 & 0.064 & 0.960 & 1.221 \\

 &  0.7 & 1000 &    1.228 & 0.115 & 0.111 & 0.946 & 1.616 & 0.140 & 0.138 & 0.958 & 1.419 \\

 & &  10,000 &  0.403 & 0.036 & 0.036 & 0.942 & 0.758 & 0.043 & 0.044 & 0.946 & 1.428 \\

 &  0.5 & 1000 &      1.732 & 0.099 & 0.097 & 0.943 & 3.186 & 0.130 & 0.136 & 0.952 & 1.699 \\ 
 
 &   & 10,000 &      0.350 & 0.031 & 0.030 & 0.946 & 0.122 & 0.040 & 0.043 & 0.945 & 1.677 \\

  \hline

\end{tabular}
 \begin{tablenotes}
     \item[1]  $MR=$ Average probability that the gold standard indicator $\Delta$ is missing at year 4 
      \item[2]  $CR=$ Average  censoring rate for the latent true event time at the end of study 
      \item[3]  $N=$ Sample size for proposed approach; if $MR > 0.0$, sample size for no auxiliary data approach is smaller because of missingness in gold standard indicator  $\Delta$.
       \item[4]  $RE=$ median relative efficiency, calculated as the median of the ratio of the estimated variance of the standard, no auxiliary data approach estimator to the estimated variance of the proposed method estimator , e.g. $\frac{Var({\hat{\beta}_{Standard}})}{Var(\hat{\beta}_{Proposed})}$
   \end{tablenotes}
    \end{threeparttable}
\end{table}

\begin{table}
\centering
      \begin{threeparttable}[t]
            \caption{Simulation results are shown for exponential failure times assuming the Cox proportional hazards model with $X\sim Normal(0.2,1)$ and $\beta=\log(1.5)$.The median percent (\%) bias, median standard errors (ASE), empirical median absolute deviation (MAD) and coverage probabilities (CP) are given for 1000 simulated data sets for the proposed method and the standard interval-censored approach that does not incorporate auxiliary data. Here, $Se = 0.80$ and $Sp = 0.90$ for the auxiliary data.}\label{table3sims}
   
      \begin{tabular}{cccccccccccccc}
\hline
\multicolumn{3}{l}{} & \multicolumn{4}{c}{Proposed}  & \multicolumn{4}{c}{No Auxiliary Data} & \multicolumn{1}{c}{ } \\ \cmidrule(r){4-7} \cmidrule{8-11}  $MR$\tnote{1}  & $CR$\tnote{2} & $N$\tnote{3}   & \% Bias & ASE & MAD & CP & \% Bias & ASE & MAD & CP & \multicolumn{1}{c}{RE\tnote{4}}  \\ 

\hline
 0.0 & 0.9 & 1000 &   -0.730 & 0.100 & 0.102 & 0.945 & -0.887 & 0.100 & 0.102 & 0.945 & 1.002 \\

 & & 10,000 &      -0.199 & 0.032 & 0.032 & 0.952 & -0.278 & 0.032 & 0.032 & 0.949 & 1.001 \\

& 0.7 & 1000 & -0.545 & 0.059 & 0.055 & 0.951 & -0.689 & 0.059 & 0.055 & 0.951 & 1.013 \\

 & & 10,000 &    0.019 & 0.018 & 0.018 & 0.950 & 0.064 & 0.019 & 0.019 & 0.949 & 1.014 \\

 & 0.5 & 1000 &       0.157 & 0.046 & 0.044 & 0.953 & 0.166 & 0.047 & 0.049 & 0.948 & 1.056 \\

 & & 10,000 &      -0.194 & 0.014 & 0.015 & 0.948 & -0.203 & 0.015 & 0.014 & 0.953 & 1.057 \\

 0.2 & 0.9 & 1000 &    -0.855 & 0.110 & 0.109 & 0.943 & -0.940 & 0.112 & 0.110 & 0.944 & 1.044 \\

 & & 10,000 &    -0.060 & 0.035 & 0.036 & 0.951 & -0.031 & 0.035 & 0.037 & 0.950 & 1.043 \\

 &  0.7 & 1000 &    -0.676 & 0.063 & 0.058 & 0.954 & -0.470 & 0.066 & 0.059 & 0.953 & 1.103 \\

 & &  10,000 &   -0.020 & 0.020 & 0.019 & 0.953 & -0.058 & 0.021 & 0.021 & 0.948 & 1.103 \\

 &  0.5 & 1000 &  0.072 & 0.049 & 0.050 & 0.954 & 0.583 & 0.053 & 0.054 & 0.939 & 1.184 \\

 &   & 10,000 &    -0.197 & 0.015 & 0.015 & 0.949 & -0.206 & 0.017 & 0.016 & 0.946 & 1.184 \\

 0.4 & 0.9 & 1000 &     -1.050 & 0.123 & 0.122 & 0.943 & 0.264 & 0.130 & 0.129 & 0.947 & 1.116 \\

 & & 10,000 &       -0.043 & 0.039 & 0.040 & 0.953 & 0.009 & 0.041 & 0.041 & 0.944 & 1.113 \\

 &  0.7 & 1000 &       -0.470 & 0.068 & 0.066 & 0.956 & -0.385 & 0.076 & 0.074 & 0.949 & 1.253 \\

 & &  10,000 &    -0.112 & 0.021 & 0.020 & 0.955 & -0.248 & 0.024 & 0.023 & 0.960 & 1.252 \\

 &  0.5 & 1000 &    0.124 & 0.052 & 0.051 & 0.949 & 0.723 & 0.061 & 0.065 & 0.937 & 1.398 \\

 &   & 10,000 &   -0.258 & 0.016 & 0.016 & 0.948 & -0.221 & 0.019 & 0.019 & 0.948 & 1.396 \\ 
  
  \hline
\end{tabular}
 \begin{tablenotes}
     \item[1]  $MR=$ Average probability that the gold standard indicator $\Delta$ is missing at year 4 
      \item[2]  $CR=$ Average  censoring rate for the latent true event time at the end of study
      \item[3]  $N=$ Sample size for proposed approach; if $MR > 0.0$, sample size for no auxiliary data approach is smaller because of missingness in gold standard indicator  $\Delta$.
      \item[4]  $RE=$ median relative efficiency, calculated as the median of the ratio of the estimated variance of the standard, no auxiliary data approach estimator to the estimated variance of the proposed method estimator , e.g. $\frac{Var({\hat{\beta}_{Standard}})}{Var(\hat{\beta}_{Proposed})}$
   \end{tablenotes}
    \end{threeparttable}
\end{table}

\begin{table}
\centering
      \begin{threeparttable}[t]
            \caption{Simulation results are shown for data simulated to be from a complex survey with exponential failure times assuming the Cox proportional hazards model with $X\sim Gamma(\textrm{shape}_s+\omega_{gs},\textrm{scale}_s+\rho_{gs})$ for an individual in block group $g$ and stratum $s$ and $\beta=\log(1.5)$. The median percent (\%) bias, median standard errors (ASE), median absolute deviation (MAD) and coverage probabilities (CP) are given for 1000 simulated data sets for the weighted proposed estimator and the weighted interval-censored approach that does not incorporate auxiliary data when both use a sandwich variance estimator to address within-cluster correlation. Here, $Se = 0.80$ and $Sp = 0.90$ for the auxiliary data.}\label{weightedtablegammaMED}
   
      \begin{tabular}{cccccccccccccc}
\hline
\multicolumn{3}{l}{} & \multicolumn{4}{c}{Proposed}  & \multicolumn{4}{c}{No Auxiliary Data} & \multicolumn{1}{c}{ } \\ \cmidrule(r){4-7} \cmidrule{8-11}  $MR$\tnote{1}  & $CR$\tnote{2} & $N$\tnote{3}   & \% Bias & ASE & MAD & CP & \% Bias & ASE & MAD & CP & \multicolumn{1}{c}{RE\tnote{4}}  \\ 

\hline
 0.0 & 0.9 & 1000 & 3.528 & 0.137 & 0.155 & 0.903 & 2.591 & 0.140 & 0.161 & 0.901 & 1.009 \\

 & & 10,000 &        2.643 & 0.044 & 0.045 & 0.923 & 1.970 & 0.044 & 0.044 & 0.937 & 1.029 \\

& 0.7 & 1000 &   4.621 & 0.098 & 0.109 & 0.920 & 5.902 & 0.102 & 0.115 & 0.910 & 1.067 \\ 
 
 & & 10,000 &     1.862 & 0.033 & 0.032 & 0.928 & 1.707 & 0.034 & 0.035 & 0.927 & 1.093 \\ 
 
 & 0.5 & 1000 &   4.714 & 0.092 & 0.100 & 0.927 & 4.735 & 0.099 & 0.107 & 0.917 & 1.167 \\

 & & 10,000 &     1.198 & 0.031 & 0.033 & 0.945 & 0.846 & 0.034 & 0.035 & 0.930 & 1.177 \\ 
 
 0.2 & 0.9 & 1000 &     3.048 & 0.146 & 0.165 & 0.902 & 0.135 & 0.153 & 0.183 & 0.912 & 1.053 \\

 & & 10,000 &    3.010 & 0.047 & 0.047 & 0.925 & 2.093 & 0.049 & 0.050 & 0.932 & 1.110 \\

 &  0.7 & 1000 &     3.695 & 0.103 & 0.113 & 0.922 & 5.268 & 0.113 & 0.132 & 0.903 & 1.225 \\

 & &  10,000 & 2.438 & 0.034 & 0.035 & 0.926 & 1.823 & 0.038 & 0.038 & 0.919 & 1.218 \\

 &  0.5 & 1000 &    3.180 & 0.097 & 0.103 & 0.923 & 3.578 & 0.110 & 0.117 & 0.931 & 1.308 \\

 &   & 10,000 &    1.243 & 0.033 & 0.034 & 0.938 & 1.120 & 0.037 & 0.038 & 0.920 & 1.354 \\

 0.4 & 0.9 & 1000 &      4.535 & 0.158 & 0.175 & 0.899 & 0.796 & 0.180 & 0.206 & 0.916 & 1.169 \\

 & & 10,000 &     2.697 & 0.050 & 0.050 & 0.931 & 1.847 & 0.057 & 0.058 & 0.930 & 1.265 \\

 &  0.7 & 1000 &   4.035 & 0.107 & 0.120 & 0.918 & 5.968 & 0.130 & 0.147 & 0.919 & 1.447 \\

 & &  10,000 &   2.805 & 0.036 & 0.037 & 0.925 & 2.259 & 0.043 & 0.047 & 0.924 & 1.455 \\

 &  0.5 & 1000 &     3.168 & 0.099 & 0.111 & 0.932 & 4.136 & 0.126 & 0.149 & 0.927 & 1.573 \\

 &   & 10,000 &   0.945 & 0.034 & 0.035 & 0.935 & 1.106 & 0.043 & 0.043 & 0.924 & 1.609 \\

  \hline
\end{tabular}
 \begin{tablenotes}
     \item[1]  $MR=$ Average probability that the gold standard indicator $\Delta$ is missing at year 4 
      \item[2]  $CR=$ Average  censoring rate for the latent true event time at the end of study
       \item[3]  $(N)=$ Average sample size for proposed approach; if $MR > 0.0$, sample size for no auxiliary data approach is smaller because of missingness in gold standard indicator  $\Delta$.
       \item[4]  $RE=$ median relative efficiency, calculated as the median of the ratio of the estimated variance of the standard, no auxiliary data approach estimator to the estimated variance of the proposed method estimator , e.g. $\frac{Var({\hat{\beta}_{Standard}})}{Var(\hat{\beta}_{Proposed})}$
   \end{tablenotes}
    \end{threeparttable}
\end{table}

\begin{table}[ht!]
\centering
      \begin{threeparttable}[t]
            \caption{Simulation results are shown for data simulated to have a similar structure to the complex survey design of HCHS/SOL, assuming exponential failure times and the Cox proportional hazards model with $\beta=\log(1.5)$. The median percent (\%) bias, median standard errors (ASE), median absolute deviation (MAD) and coverage probabilities (CP) are given for 1000 simulated data sets for the proposed estimator and the interval-censored approach that does not incorporate auxiliary data when both apply regression calibration to address covariate error. Variance estimation is performed using the resampling-based multiple imputation procedure of Baldoni et al. (2021).}\label{HCHSSIMS}
   
      \begin{tabular}{cccccccccccccc}
\hline
\multicolumn{4}{c}{Proposed}  & \multicolumn{4}{c}{No Auxiliary Data} & \multicolumn{1}{c}{ } \\ \cmidrule(r){1-4} \cmidrule{5-8}    \% Bias & ASE & MAD & CP & \% Bias & ASE & MAD & CP & \multicolumn{1}{c}{RE\tnote{1}}  \\ 

\hline

   -0.859 & 0.203 & 0.189 & 0.949 & -1.470 & 0.244 & 0.237 & 0.950 & 1.442 \\ 
  \hline
\end{tabular}
 \begin{tablenotes}
     \item[1]    $RE=$ median relative efficiency, calculated as the median of the ratio of the estimated variance of the standard, no auxiliary data approach estimator to the estimated variance of the proposed method estimator , e.g. $\frac{Var({\hat{\beta}_{Standard}})}{Var(\hat{\beta}_{Proposed})}$
   \end{tablenotes}
    \end{threeparttable}
\end{table}

\clearpage 

\begin{table}[ht]
\centering
      \begin{threeparttable}[t]
            \caption{Type I error results for $\beta=0$ are given for 1000 simulated data sets for the proposed method when data are simulated using exponential failure times and assuming the Cox proportional hazards model with $X\sim Gamma(0.2,1)$. Here, $Se = 0.80$ and $Sp = 0.90$ for the auxiliary data.}\label{t1errortable}
   
      \begin{tabular}{cccccccccccccc}
      \hline
\multicolumn{1}{c}{}  & \multicolumn{1}{c}{} & \multicolumn{3}{c}{Type I Error Rate} \\ \cmidrule(r){3-5}    $CR$\tnote{1} & $N$\tnote{2} & $MR \tnote{3}=0.0$ & $MR=0.2$ & $MR=0.4$ \\ 

\hline
 0.9 & 1000 &  0.045 & 0.033  & 0.049  \\

 & 10,000 &  0.056 & 0.065 &    0.054 \\

 0.7 & 1000 &   0.043  & 0.049 & 0.061 \\

  & 10,000 &     0.047 &  0.045 & 0.048 \\

  0.5 & 1000 &   0.050 & 0.049 & 0.057  \\

 & 10,000 &    0.051 &   0.056 &  0.051 \\

 \hline

\end{tabular}
 \begin{tablenotes}
     \item[1] $CR=$ Average  censoring rate for the latent true event time at the end of study
      \item[2]  $N=$ Sample size for proposed approach; if $MR > 0.0$, sample size for no auxiliary data approach is smaller because of missingness in gold standard indicator  $\Delta$.
     \item[3]   $MR=$ Average probability that the gold standard indicator $\Delta$ is missing at year 4 
   \end{tablenotes}
    \end{threeparttable}
\end{table}

\clearpage 

\begin{table}
\centering
 \begin{threeparttable}[t]
\caption {HCHS/SOL Data Analysis on a random subset $(N=8,200)$ of study participants using baseline sensitivity ($Se = 0.61$) and specificity ($Sp = 0.98$) values. Hazard Ratio (HR) and 95\% confidence interval (CI) estimates of incident diabetes for a 20\% increase in consumption of energy (kcal/d), protein (g/d), and protein density (\% energy from protein/d) based on the proposed estimator and the interval-censored approach that does not incorporate auxiliary data.} \label{tableHR} 

\centering
\begin{tabular}{llll}
\hline
\multicolumn{1}{c}{}  & \multicolumn{2}{c}{HR (95\% CI)} & \multicolumn{1}{c}{ } \\ \cmidrule(r){2-3}  \multicolumn{1}{c}{Model\tnote{1}} & \multicolumn{1}{c}{Proposed} & \multicolumn{1}{c}{No Auxiliary Data}  & \multicolumn{1}{c}{RE\tnote{2}} \\  
  \hline
 Energy (kcal/d)   &   1.20 (0.47, 3.11) & 1.20 (0.41, 3.82) & 1.27 \\

   Protein (g/d)   &    1.30 (0.82, 2.06) & 1.37 (0.74, 2.51) & 1.74 \\

  Protein Density   &    1.01 (1.00, 1.02) & 1.01 (1.00, 1.03) & 1.63 \\

   \hline
\end{tabular}
 \begin{tablenotes}
     \item[1] Each model is adjusted for potential confounders including age, body mass index (BMI), sex, Hispanic/Latino background, language preference, education, income, and smoking status.
           \item[2]  $RE=$ relative efficiency, calculated as the ratio of the estimated variance of the standard, no auxiliary data approach estimator to the estimated variance of the proposed method estimator , e.g. $\frac{Var({\hat{\beta}_{Standard}})}{Var(\hat{\beta}_{Proposed})}$

   \end{tablenotes}
    \end{threeparttable}%
\end{table}

\clearpage 

\newpage

\begin{center}
{\Large \textbf{Supplementary Materials for ``An Augmented Likelihood Approach for the Discrete Proportional Hazards Model Using Auxiliary and Validated Outcome Data – with Application to the HCHS/SOL Study"} }
\end{center}

\begin{center}
\normalsize{Lillian A. Boe$^{1,*}$ and 
Pamela A. Shaw$^{2}$\\
$^{1}$Department of Biostatistics, Epidemiology, and Informatics, \\ University of Pennsylvania Perelman School of Medicine, \\  Philadelphia, PA 19104\\
$^{2}$Kaiser Permanente Washington Health Research Institute, \\ Seattle, Washington 98101 \\

\textit{*email}: boel@pennmedicine.upenn.edu}

\end{center}

\begin{singlespace}
\section*{Contents}
\begin{itemize}
    \item  \hyperref[rcodesample]{S1: R Code with sample data analysis} 
        \item \hyperref[cijlikelihood]{S2: Details of $C_{ij}$ term in the likelihood}
    \item \hyperref[theory]{S3: Regularity conditions for asymptotic normality}
    \begin{itemize}
         \item \hyperref[randsamptheory]{S3.1: Proposed estimator for random sample}
  \item \hyperref[complexsurvtheory]{S3.2: Proposed estimator for complex survey design}
    \item \hyperref[complexsurvRC]{S3.3: Proposed estimator for complex survey design and regression calibration}
  \end{itemize}
\item \hyperref[HCHSSOLSUPPDETAILS]{S4: Supplemental details for HCHS/SOL data example}
\item \hyperref[tablesuppNotErrProne]{Supplemental tables}
\begin{itemize}
    \item \hyperref[tablesuppNotErrProne]{Table S1}
        \item \hyperref[supptableSESP]{Table S2}
    \item \hyperref[supptablelog3]{Table S3}
    \item \hyperref[weightedtablenormalMED]{Table S4}
    \item \hyperref[tableHRSensAnalysis]{Table S5}
\end{itemize}
\end{itemize}
\end{singlespace}

\setcounter{equation}{0}
\renewcommand{\theequation}{S\arabic{equation}}

\setcounter{section}{0}
\renewcommand{\thesection}{S\arabic{section}}

\setcounter{table}{0}
\renewcommand{\thetable}{S\arabic{table}}

\section{R Code with sample data analysis}\label{rcodesample}

We now provide R code that illustrates how to apply the proposed method and the standard, no auxiliary data method to a simulated data set. The simulated data mimics the complex survey design of HCHS/SOL that includes unequal probability sampling, stratification, and clustering. To mimic the measurement error of dietary factors in HCHS/SOL, we simulate an error-prone covariate ($X^{*}$) and assume that we additionally have 2 error-free continuous covariates, such as body mass index and age ($Z_1$ and $Z_2$). The auxiliary data outcome variable is recorded at 8 time points, while the gold standard outcome variable is recorded at year 4. The sensitivity and specificity of the error-prone, auxiliary data outcome are assumed to be $0.61$ and $0.98$, respectively. This data set is provided on GitHub at \url{https://github.com/lboe23/AugmentedLikelihood} with the file name SampleData.RData.

We begin by loading in the functions required to calculate our log-likelihood and gradient. These functions are available on the GitHub site above in the file titled ``PROPOSED\_AUGMENTEDLIKELIHOOD\_FUNCTIONS.R." This file contains two functions, (1) log\_like\_proposed() which calculates the log-likelihood for the proposed method and (2) gradient\_proposed() which calculates the gradient/estimating function for the proposed method. Both functions require a specification of the function purpose, where the options are ``SUM" or ``INDIVIDUAL." SUM returns the sum of the log-likelihood or gradient contributions, respectively, while individual returns a vector/matrix of each person's contributions. We also use the Rcpp function cmat() available in the file ``RcppFunc.cpp" from GitHub. Additionally, we use the Rcpp functions dmat() and getrids() developed by \citet{gu2015semiparametric} that can be found in the icensmis package on Cran or on GitHub at \url{https://github.com/XiangdongGu/icensmis/blob/master/src/dataproc.cpp}. Below is code that loads all of the required functions:

\begin{lstlisting}
source(`PROPOSED_AUGMENTEDLIKELIHOOD_FUNCTIONS.R')
Rcpp::sourceCpp(`RcppFunc.cpp')
Rcpp::sourceCpp(`dataproc.cpp')
\end{lstlisting}

Now we assign the sensitivity ($Se$) and specificity ($Sp$) values for the auxiliary data. We assume these are known, fixed constants in our analysis. We will also allow for a proportion of the gold (reference) standard event status variables to be missing, and we fix this missingness rate to be 20\% for this analysis.

\begin{lstlisting}
    sensitivity<-0.61
    specificity<-0.98
    prop_m<-0.20
\end{lstlisting}

Now, we load in sample simulated data. The data we input is in wide form, with one row per subject and each auxiliary data event indicator in a separate column.

\begin{lstlisting}
     load(file=paste0('SampleData.RData'))
     N<-dim(samp)[1]
\end{lstlisting}

We now fit the calibration model. Later, for variance estimation in the presence of regression calibration and a complex survey design, we will use the parametric multiple imputation procedure proposed by \citet{baldoni2021use}. To do so, we need to save off the estimated calibration model regression coefficients, corresponding estimated covariance matrix, and the design matrix. Finally, we will create our predicted values (xhat) from regression calibration using the ``predict" statement.

\begin{lstlisting}
    samp.solnas <- samp[(solnas==T),]
    lm.lsodi <- glm(xstarstar ~  xstar+z1+z2,data=samp.solnas) 
    x.lsodi <- model.matrix(lm.lsodi) #X
    xtx.lsodi <- t(x.lsodi)%*%x.lsodi #X`X`
    ixtx.lsodi <- solve(xtx.lsodi) #(X`X)^-1
    samp[,xhat := predict(lm.lsodi,newdata=samp,'response')]
\end{lstlisting}

We now convert the data to long form, where each row represents one time point and each subject has multiple rows. Recall that each subject in our simulated data set has 8 visits. Then, we are going to sort the data by subject ID.

\begin{lstlisting}
  samp_long1<-reshape(data = samp, idvar = "ID", varying = 
  list(true_result=
  c("true_result_1", "true_result_2","true_result_3","true_result_4",
  "true_result_5","true_result_6","true_result_7","true_result_8"), 
  result=c("result_1","result_2","result_3", "result_4","result_5",
         "result_6", "result_7","result_8")), direction="long",
                      v.names = c("true_result","result"),sep="_")

#order long dataset by each subject's ID
  samp_long <- samp_long1[order(samp_long1$ID),]
\end{lstlisting}

Next, we create a data set with only one row per subject using the duplicated function and applying it to the ID variable. Then, using the data with one row per subject, we save the vector of sampling weights that will be used in the weighted analysis. Additionally, we create a keep statement and apply a function called ``after\_first\_pos" to removes all auxiliary data values of ``1" after the first positive from the data in long form. 

\begin{lstlisting}
  GS_data<-samp_long[!duplicated(samp_long$ID),c("ID","GS_vis4")]

#Save vector of weights for this dataset
  weights<-as.numeric(unlist(samp_long[!duplicated(samp_long$ID),
  c("bghhsub_s2")]))

keep<-unlist(tapply(samp_long$result,samp_long$ID,after_first_pos))
 datafinal_1<-samp_long[keep,]
\end{lstlisting}

Suppose we want to simulate missingness in the gold (reference) standard indicator variable, $\Delta_i$. To do so, we first set a seed so that our results are reproducible. Then, we generate $N$ variables called $mcar$ from a Uniform(0,1) distribution and let $\Delta_i$ be missing for each subject if $mcar<MR$. 
\begin{lstlisting}
  set.seed(2548)
  mcar<-runif(N,0,1)
  GS_data$GS_vis4_mis<-ifelse(mcar<prop_m,NA,GS_data$GS_vis4)
\end{lstlisting}

Next, we create two datasets using the merge function: datafinal, which is the data in long form with one row per visit, and datafinal\_GS, which has one row per person for the standard, no auxiliary data analysis. 

\begin{lstlisting}
  datafinal<-merge(datafinal_1,GS_data,by="ID")
  datafinal_GS<-merge(samp_long,GS_data[,c("ID","GS_vis4_mis")],by="ID")
\end{lstlisting}

Now, we write down the formula for our outcome model including the error-prone auxiliary data outcome and three covariates. Recall that we used regression calibration to correct for error in one covariate, so our model includes the predicted value xhat and two precisely recorded covariates, z1 and z2.

\begin{lstlisting}
  formula=result~xhat+z1+z2
\end{lstlisting}

We will now make sure our data is ordered properly before we begin calculating sum of the likelihood components.

\begin{lstlisting}
  id <- eval(substitute(datafinal$ID), datafinal, parent.frame())
  time <- eval(substitute(datafinal$time), datafinal, parent.frame())
  result <- eval(substitute(result), datafinal, parent.frame())
  ord <- order(id, time)
  if (is.unsorted(ord)) {
    id <- id[ord]
    time <- time[ord]
    result <- result[ord]
    datafinal <- datafinal[ord, ]}
  utime <- sort(unique(time))
  timen0 <- (time != 0)
\end{lstlisting}

Next, we will calculate the D matrix and C matrix for our log-likelihood using the Rcpp functions. Additionally, we assign J, the number of auxiliary data visit times.

\begin{lstlisting}
  Dm <- dmat(id[timen0], time[timen0], result[timen0], sensitivity,
             specificity, 1)
  Cm <- cmat(id[timen0], time[timen0], result[timen0], sensitivity,
             specificity, 1)
  J <- ncol(Dm) - 1
\end{lstlisting}

In these next steps, we create our covariate matrix with one column per covariate (Xmat). Intially,  Xmat will be in long form, with one row per visit. We also assign nbeta (the number of covariates/regression parameters) and uid, a unique indicator for each person's ID. Finally, we redefine our covariate matrix to have just one row per subject.

\begin{lstlisting}
  Xmat <- model.matrix(formula, data = datafinal)[, -1, drop = F]
  beta.nm <- colnames(Xmat)
  nbeta <- ncol(Xmat)
  uid <- getrids(id, N)
  Xmat <- Xmat[uid, , drop = F]
\end{lstlisting}

Now, create a unique vector (GSdelta) with only one row per person indicating whether each person had the gold (reference) standard indicator available or not. This will be used to calculate the proposed log-likelihood contribution for each subject based on whether GSdelta=NA, 0 or 1. Additionally, we create the vector of observation times at which the gold (reference) standard is recorded, called GSVis. Lastly, we create a vector of 1's called ``noweights" which will be used to fit the proposed estimator in the unweighted analysis.

\begin{lstlisting}
  GSdelta <- datafinal[uid,"GS_vis4_mis"]
  GSVis<-rep(4,N)
  noweights<-rep(1,N)
\end{lstlisting}

We now finalize the data set for the standard, no auxiliary data analysis using the unique IDs only such that data has $N$ rows. This is the final dataset for standard, no auxiliary data analysis analysis that omits anyone who is missing the gold standard.

\begin{lstlisting}
  IC_data<-datafinal_GS[!duplicated(datafinal_GS$ID),c("result",
  "true_result","GS_vis4","GS_vis4_mis","BGid", "strat", 
  "bghhsub_s2","xstar","xhat","z1","z2")]
  IC_GS_datafinal<-IC_data[complete.cases(IC_data$GS_vis4_mis),]
\end{lstlisting}

Now, we create starting values for our survival parameters. First, to avoid maximization problems due to the ordered constraint of the survival parameters, we re-parameterize these in terms of a log-log transformation of survival function for $S_2$, and a change in log-log of the survival function for all other parameters $S_3 \ldots S_{J+1}$. We consider initial values of 0.1 for our survival parameters, then transform these based on this re-parameterization. We also define lower and upper bounds for our survival parameters. Our lower bound is infinity for the first re-parameterized survival function and 0 for the subsequent $J-1$ terms. Our upper bound is infinity for all terms. Finally, we create a vector $parmi$ consisting of a starting value for $beta$, and starting values for re-parameterized survival parameters.

\begin{lstlisting}
  initsurv <- 0.1
  lami <- log(-log(seq(1, initsurv, length.out = J + 1)[-1]))
  lami <- c(lami[1], diff(lami))
  tosurv <- function(x) exp(-exp(cumsum(x)))
  lowlam <- c(-Inf, rep(0, J - 1))
  lowerLBFGS <- c(rep(-Inf, nbeta),lowlam)
  upperLBFGS <- c(rep(Inf, nbeta+J))
  parmi <- c(rep(0.5,nbeta),lami)
\end{lstlisting}

Now, we create survey designs using survey package for the two models: the proposed method, with all $N$ subjects, and the standard, no auxiliary data analysis model which excludes subjects missing the gold standard variable. 

\begin{lstlisting}
samp_design_reg = svydesign(id=~BGid, strata=~strat,       
    weights=~bghhsub_s2, data=IC_GS_datafinal)
samp_design_reg_complete = svydesign(id=~BGid, strata=~strat, 
    weights=~bghhsub_s2, data=IC_data)
\end{lstlisting}

Finally, we fit the proposed method and the standard, no auxiliary data approach with the weights from the survey design. One may consider the functions \texttt{optim()} or \texttt{nlminb()} for maximization of the log-likelihood in R.

\begin{lstlisting}
proposed_fit_data_weight<-optim(par=parmi, fn=log_like_proposed,
    gr=gradient_proposed,lower = lowerLBFGS,upper=upperLBFGS,
    method = "L-BFGS-B",N,J,nbeta,Dm,Cm,Xmat,GSdelta,GSVis,
    weights=weights,purpose="SUM",hessian=TRUE)
inverted_hessian<-solve(proposed_fit_data_weight$hessian)

proposed_fit_data_weight_GS<-svyglm(GS_vis4~xhat+z1+z2,
    family=quasibinomial(link="cloglog"),data=IC_GS_datafinal,
    design=samp_design_reg)
\end{lstlisting}

Next, we calculate a matrix of estimating equation contributions for all individuals. We want the unweighted matrix, because the function \texttt{svytotal()} adds the weights later. Then, we compute the influence functions by multiplying this matrix by the inverse of the hessian matrix. Finally, we obtain design-based standard errors using the variance approach of \citet{binder1983variances} and functions from the survey package by providing the influence function and survey design to \texttt{vcov(svytotal())}  \citep{lumley2011complex}.
\begin{lstlisting}
U_prop1<-gradient_proposed(proposed_fit_data_weight$par,N,J,nbeta,
    Dm,Cm,Xmat,GSdelta,GSVis,weights=noweights,purpose="INDIVIDUAL")
infl1<- U_prop1%*%inverted_hessian
mySandVar<- vcov(svytotal(infl1,samp_design_reg_complete,
    influence=TRUE))
\end{lstlisting}

We now save the estimated parameters, including the estimated regression coefficients and corresponding standard error estimates.

\begin{lstlisting}
beta1est_aux_w<-  proposed_fit_data_weight$par[1]
sandVar<-  sqrt(diag(mySandVar))[1]
fitsum_truth4Year<-summary(proposed_fit_data_weight_GS)
beta1est_truth4Year<-fitsum_truth4Year$coefficients["xhat",1]
beta1se_truth4Year<-fitsum_truth4Year$coefficients["xhat",2]
\end{lstlisting}

The final step of our analysis is to use the parametric multiple imputation procedure of \citet{baldoni2021use} to compute the variance that accounts for the extra uncertainty added by the calibration model step. Code for implementing this procedure is available on GitHub at \url{https://github.com/plbaldoni/HCHSsim} and involves sampling the calibration model coefficients using their asymptotic parametric distribution $M$ times, resulting in $M$ sets of calibration coefficients. Then, we are able to estimate $M$ predicted covariate (xhat) values, $\hat{X}_i^{(m)}$. Finally, we fit the outcome models each $M$ times, once for each of the newly predicted intake values $\hat{X}_i^{(m)}$ from multiple imputation. We chose $M=25$ imputations for our analysis. We present full code that applies the procedure of \citet{baldoni2021use} to the proposed estimator fit to simulated data on GitHub at \url{https://github.com/lboe23/AugmentedLikelihood} in the code file ``Sample\_Data\_Analysis\_Final.R" This full code for running the multiple imputation variance procedure involves running all of the above code in a multiple imputation loop, and thus to avoid redundancy we have chosen to omit it here. On GitHub, we have also provided the output from applying this parametric multiple imputation procedure, titled ``SampleAnalysis\_MIVarianceResults.RData". Below, we load in this output so that we can obtain our final variance calculations.

\begin{lstlisting}
load(file=paste0('SampleAnalysis_MIVarianceResults.RData'))
output_mi<-as.data.frame(matrix(unlist(list_mi), ncol=5, byrow=T))
colnames(output_mi)<-c("Imp","beta_proposed_mi","se_proposed_mi",
"beta_standard_mi","se_standard_mi")
\end{lstlisting}

Now, let's use this output to calculate the final estimate of the variance of the regression coefficient $\hat{\beta}$ from each model as: $\hat{V}^*=\frac{1}{M}\sum_{m=1}^M \hat{V}^{(m)}+\frac{1}{M-1}\sum_{m=1}^M\Big(\hat{\beta}^{(m)}-\bar{\hat{\beta}}\Big)^2$, where $\bar{\hat{\beta}}=\frac{1}{M}\sum_{m=1}^M\hat{\beta}^{(m)}$ and $\hat{\beta}^{(m)}$ and $\hat{V}^{(m)}$ represent the estimated regression coefficient and its estimated variance, respectively, using the $m$-th completed data set with $m=1,\ldots,M$. Recall that this formula for computing variance estimates accounting for regression calibration were described in section 2.3.2 from the main manuscript. Following \citet{baldoni2021use}, we use robust estimators for the mean and standard deviation in this equation, specifically the median and the median absolute deviation:

\begin{lstlisting}
MI_Var<-function(V,betas){
      var<-(median(V)+(mad(betas)^2))
      return(sqrt(var))
}

se_proposed_final<-MI_Var((output_mi$se_proposed_mi)^2,
output_mi$beta_proposed_mi)

se_standard_final<-MI_Var((output_mi$beta_standard_mi)^2,
output_mi$beta_standard_mi)
\end{lstlisting}

Now, we can use our estimated regression coefficients and their corresponding standard errors to construct a table with estimated hazard ratios (HR) and 95\% confidence intervals (CI) associated with a 20\% increase in consumption. We will also include relative efficiency calculations in our table, computed as the ratio of the estimated variance of the standard, no auxiliary data approach estimator to the estimated variance of the proposed method estimator, e.g. $\frac{Var({\hat{\beta}_{Standard}})}{Var(\hat{\beta}_{Proposed})}$:

\begin{lstlisting}
myfinaltable<-cbind(paste(round(exp(beta1est_aux_w*log(1.2)),2),"(",
    round(exp(beta1est_aux_w-1.96*se_proposed_final)^log(1.2),2),",",
    round(exp(beta1est_aux_w+1.96*se_proposed_final)^log(1.2),2),")"),
    paste(round(exp(beta1est_truth4Year*log(1.2)),2),
    "(",round(exp(beta1est_truth4Year-1.96*se_standard_final)^log(1.2),2),
    ",",round(exp(beta1est_truth4Year+1.96*se_standard_final)^log(1.2),2),
    ")"),round((se_standard_final^2)/(se_proposed_final^2),2))
\end{lstlisting}

Below are our final results from the table we just constructed:

\begin{center}
\begin{tabular}{lll}
\hline \multicolumn{2}{c}{HR (95\% CI)} & \multicolumn{1}{c}{ } \\ \cmidrule(r){1-2}   \multicolumn{1}{c}{Proposed} & \multicolumn{1}{c}{No Auxiliary Data}  & \multicolumn{1}{c}{RE\tnote{2}} \\  
  \hline
1.07 (0.99, 1.16) & 1.08 (0.98, 1.19) & 1.51 \\ 
\hline
\end{tabular}
\end{center}

\section{Details of $C_{ij}$ term in the likelihood}\label{cijlikelihood}

In Section 2.1 of the main text, we introduce the likelihood contributions for all individuals in terms of $C_{ij}=\left[\prod_{l=1}^{n_i}P(Y^*_{il}|\tau_{j-1}<T_i\leq \tau_j,T_{il}^*,\Delta_i)\right]$, which is simply a function of the sensitivity $(Se)$ and specificity $(Sp)$ of the auxiliary data. Recall that we assume constant and known sensitivity $(Se)$ and specificity $(Sp)$, defined as $Se=\Pr(Y^*_{il}=1|\tau_{j-1}<T_i\leq\tau_j,T_l^*\geq\tau_j)$ and $Sp=\Pr(Y^*_{il}=0|\tau_{j-1}<T_i\leq\tau_j,T_l^*\leq\tau_{j-1})$. It is then straightforward to see that $1-Se=\Pr(Y^*_{il}=0|\tau_{j-1}<T_i\leq\tau_j,T_l^*\geq\tau_j)$ and $1-Sp=\Pr(Y^*_{il}=1|\tau_{j-1}<T_i\leq\tau_j,T_l^*\leq\tau_{j-1})$In this section, we provide the general form of the $C_{ij}$ terms for the case of no missed visits. The formula introduced below could be modified to allow for missed visits by summing up all terms $\theta_j=\Pr(\tau_{j-1}<T_i\leq\tau_j)$ across the $(\tau_{j-1},\tau_j]$ defining a subject's observational interval. For the case of $J$ total visits for all $N$ subjects, the $C_{ij}$ terms take the following form:
\begin{align*}
    \begin{matrix}  C_{i1}=Se^{\sum_{j=1}^{n_i}Y_{ij}^*}(1-Se)^{\sum_{j=1}^{n_i}(1-Y_{ij}^*)},\\  C_{i2}=Sp^{(1-Y_{i1})}(1-Sp)^{Y_{i1}}Se^{\sum_{j=2}^{n_i}Y_{ij}^*}(1-Se)^{\sum_{j=2}^{n_i}(1-Y_{ij}^*)}, \\ 
 ... \\
    C_{i(J+1)}=Sp^{\sum_{j=1}^{n_i}(1-Y_{ij}^*)}(1-Sp)^{\sum_{j=1}^{n_i}Y_{ij}^*}. \end{matrix}
\end{align*}

As an example, consider subject $i=1$ who has observed auxiliary data vector $\mathbf{Y^*_i}=[0, 0, 0, 1]$ corresponding to annual visit time vector $\mathbf{T^*_i}=[1, 2, 3, 4]$. Suppose the visit times among all $N$ subjects are also $[\tau_0=0, \tau_1=1, \tau_2=2, \tau_3=3, \tau_4=4, \tau_{5}=\infty]$. Then, for subject $i$ with $n_i=J=4$ and $j=1$, we have:

\begin{eqnarray*}
    C_{11}&=&\prod_{l=1}^{4}P(Y^*_{1l}|\tau_{0}<T_1\leq \tau_1,T_{1l}^*,\Delta_1) \\
     C_{11}&=&P(Y^*_{11}|\tau_{0}<T_1\leq \tau_1,T_{11}^*,\Delta_1) \times P(Y^*_{12}|\tau_{0}<T_1\leq \tau_1,T_{12}^*,\Delta_1) \times  \\ && P(Y^*_{13}|\tau_{0}<T_1\leq \tau_1,T_{13}^*,\Delta_1) \times P(Y^*_{14}|\tau_{0}<T_1\leq \tau_1,T_{14}^*,\Delta_1)\\
      C_{11}&=&P(Y^*_{11}=0|\tau_{0}<T_1\leq \tau_1,T_{11}^* \geq \tau_1,\Delta_1) \times P(Y^*_{12}=0|\tau_{0}<T_1\leq \tau_1,T_{12}^*\geq \tau_1,\Delta_1) \times  \\ && P(Y^*_{13}=0|\tau_{0}<T_1\leq \tau_1,T_{13}^*\geq \tau_1,\Delta_1) \times P(Y^*_{14}=1|\tau_{0}<T_1\leq \tau_1,T_{14}^*\geq \tau_1,\Delta_1)\\
      C_{11}&=& (1-Se) \times (1-Se) \times (1-Se) \times Se\\
\end{eqnarray*}

Then, following a similar procedure for $j=2,\ldots,5$, we see that for subject $1$,

\begin{align*}
    \begin{matrix}  C_{11}=Se(1-Se)^3,\\  C_{12}=SpSe(1-Se)^2, \\ 
 C_{13}=Sp^2 Se(1-Se),\\
 C_{14}=Sp^3 Se\\
 C_{15}=Sp^3 (1-Sp). \end{matrix}
\end{align*}

\section{Regularity conditions for asymptotic normality}\label{theory}

We now provide sufficient regularity conditions for the proposed estimator to be asymptotically normal and achieve a $\sqrt{N}$-convergence rate. We assume the following throughout this section: (1) $\{T_i,\Delta_i,Y_i^*,T_i^*,M_i,X_i,X_i^*,Z_i\}$ is a vector of independent and identically distributed random variables for $i=1,...,N$, where $N$ is the number of subjects in the main study data; (2) $T_i$ is the latent, unobserved continuous failure time of interest for subject $i$; (3) the proportional hazards model holds for the latent true event time, such that $S(t)=S_0(t)^{\exp(x'\beta)}$; (4) the auxiliary outcome status ($Y_i^*$) is observed at $n_i$ follow-up times $T_i^*=(t_{i1},...,t_{in_i})$   that are a subset of J+1 possible observation times  satisfying $0=\tau_0<\tau_1<\tau_2<...<\tau_J <\tau_{J+1} = \infty$; and (3) $\tau_{V_i}$ is the observation time for the gold standard event status variable $\Delta_i$, where $\tau_{V_i} \in \{\tau_0,\tau_1,\tau_2,...,\tau_J\}$; (5) $\tau_{V_i}$ is the gold standard assessment time for individual $i$, where $\tau_{V_i} \in \{\tau_0,\tau_1,\tau_2,...,\tau_J\}$ and $\Delta_i$ is the corresponding gold standard event status variable; (6) the binary variable $M_i \in \{0,1\}$ indicates whether $\Delta_i$ is missing; (7) $X_i$ is the $p$-dimensional true covariate vector of interest with corresponding error-prone vector $X_i^*$, while $Z_i$ is the additionally observed $q$-dimensional vector of error-free covariates, where all covariates are random variables with finite variance. In the main text, we define $\theta_j=\Pr(\tau_{j-1}<T_i\leq\tau_j)$ and $S_j=\sum_{h=j}^{J+1}\theta_h=\Pr(T>\tau_{j-1})$ for $j=1,...,J+1$. Additionally, we require that $1=S_1>S_2>...>S_{J+1}>0$, ensuring that $0 < \theta_j < 1$ for $j=1,\ldots,J$.

\subsection{Proposed estimator for random sample}\label{randsamptheory}

First, we consider the case where the data are assumed to be a simple random sample from the population and the covariates of interest are recorded precisely (i.e. error-free).  The log-likelihood function $l(\psi)=l(T_i,\Delta_i,Y_i^*,T_i^*,M_i,X_i;\psi)$ is defined in section 2.1, equation 2.4 from the main text as:

\begin{eqnarray}\label{loglikelihoodallNtheorysupp}
l(\psi)= l(S,\beta) = \sum_{i=1}^N  l_i(S,\beta) 
&=&  \sum_{i=1}^N \Bigg[ (1-M_i)\Delta_i \log \Bigg( \sum_{j=1}^{V_i} D_{ij} (S_j)^{\exp(x_i'\beta)} \Bigg) + \nonumber \\ &&  (1-M_i)(1-\Delta_i) \log \Bigg( \sum_{j=V_i+1}^{J+1} D_{ij} (S_j)^{\exp(x_i'\beta)}  \Bigg) 
+ \nonumber \\ && M_i \log \Bigg( \sum_{j=1}^{J+1} D_{ij} (S_j)^{\exp(x_i'\beta)}  \Bigg)\Bigg].
\end{eqnarray}

\noindent where $\mathbf{\psi=[\beta,S]}$ and $\mathbf{S} = (S_1,S_2,...,S_{J+1})'$. Recall that we define the score function as $U_i({\psi})=\frac{\partial l_i(\psi) }{\partial \psi}$. The proposed estimator in this setting, $\hat{\psi}$, is found by solving the score equation $\sum_{i=1}^N U_i({\psi})=0$. Let $\psi^0$ be the true vector of regression parameters of interest that solves $E\left[\frac{\partial l(\psi)}{\partial \psi}\right]=E\left[\sum_{i=1}^N U_i({\psi})\right]=0$. Now, further assume that the log-likelihood $l(\psi)=l(T_i,\Delta_i,Y_i^*,T_i^*,M_i,X_i;\psi)$ is twice continuously differentiable with respect to $\psi$ such that there exists an invertible Hessian matrix \citep{foutz1977unique}. We additionally assume the regularity conditions made by \citet{foutz1977unique} for establishing consistency and uniqueness of our estimator. Then, appealing to standard maximum likelihood estimation (MLE) theory, with probability going to one as $N \rightarrow \infty$, $\hat{\psi}$ is a unique solution to the likelihood equations that is consistent for $\psi^0$ and asymptotically normal \citep{boos2013essential}. Specifically, one has:
\begin{equation}
    \sqrt{N}(\hat{\psi}-\psi^0) \xrightarrow{d} \mathcal{N}(0, {\text{V}}({\psi}^0)),
\end{equation}

\noindent where ${\text{V}}({\psi}^0)$ is the Fisher information matrix, denoted by $=I(\psi^0)^{-1}$.

\subsection{Proposed estimator for complex survey design}\label{complexsurvtheory}

We now state the additional regularity conditions needed for the proposed method estimator to accommodate data from a complex survey sampling design by using a weighted log-likelihood function and a sandwich variance estimator to address within-cluster correlation.

Recall from section 2.2 that a sample of $N$ subjects is drawn from a population of size $N_{POP}$ resulting in the sampling probability $\pi_i$. Following \citet{lumley2017fitting}, we assume that $N \rightarrow \infty$ and $\frac{N}{N_{POP}} \rightarrow p\in (0,1)$. Additionally assume that $\pi_i$ is known for the subjects in the sample and is bounded away from 0. The weighted log-likelihood equation  of the main text is written as follows: $l_{\pi}(S,\beta) = \sum_{i=1}^N \frac{1}{\pi_i}  l_i(S,\beta) = \sum_{i=1}^N  \widecheck{l}_i(S,\beta).$  The weighted proposed  estimator $\hat{\psi}_{\pi}$ may be found by solving the weighted score equation, $\sum_{i=1}^N \widecheck{U}_{i}({\psi}_{\pi})=\sum_{i=1}^N \frac{1}{\pi_i} U_{i}({\psi}_{\pi})=0$. Then, as before, $\psi_{\pi}^0$ is the solution to $E\left[\frac{\partial l_{\pi}(\psi_{\pi})}{\partial \psi_{\pi}}\right]=E\left[\sum_{i=1}^N \widecheck{U}_i({\psi}_{\pi})\right]=0$. Following the same logic applied for the random sample case and since $\hat{\psi}_{\pi}$ is also a maximum likelihood estimator, we have:

\begin{equation}
    \sqrt{N}(\hat{\psi}_{\pi}-\psi_{\pi}^0) \xrightarrow{d} \mathcal{N}(0, {\text{V}}({\psi}^0_{\pi})),
\end{equation}

\noindent where ${\text{V}}({\psi}^0_{\pi})$ can be approximated using the implicit differentiation method of \citet{binder1983variances}. To use this variance estimation approach, assume that $\sum_{i=1}^N \widecheck{U}_{i}(\psi_{\pi})$ is suitably smooth such that $\hat{\psi}_{\pi}$ can be implicitly defined as a function of $\sum_{i=1}^N \widecheck{U}_{i}(\psi_{\pi})$. Further assume that the derivative matrix for $\widecheck{U}_{i}$ is full rank, invertible, and a continuous function of $\psi_{\pi}$. Then, we can use a Taylor expansion of $U_{i}$ at $\hat{\psi}_{\pi}=\psi^0_{\pi}$ to arrive at the following estimator for the asymptotic variance of $\hat{\psi}_{\pi}$: $\hat{\text{V}}[\hat{\psi}_{\pi}] \approx \left(\sum_{i=1}^N \frac{\partial \widecheck{U}_{i} (\hat{\psi}_{\pi})}{\partial \psi_{\pi}}\right)^{-1} \hat{\text{cov}}\left[\sum_{i=1}^N \widecheck{U}_{i} (\hat{\psi}_{\pi}) \right] \left(\sum_{i=1}^N \frac{\partial \widecheck{U}_{i} (\hat{\psi}_{\pi})}{\partial \psi_{\pi}}\right)^{-1}$. Details and theoretical justification are provided in \citet{binder1983variances}.

\subsection{Proposed estimator for complex survey design and regression calibration}\label{complexsurvRC}
 
First assume that the error-free covariates $X_i$ and $Z_i$ are available for all sampled individuals, such that our log-likelihood $l_{\pi,X,Z}(\psi)=l(T_i,\Delta_i,Y_i^*,T_i^*,M_i,X_i, Z_i;\psi)$ takes the following form:

 \begin{eqnarray}\label{loglikelihoodallNxz}
l_{\pi,X,Z}(\psi) = \sum_{i=1}^N  \frac{1}{\pi_i}  l_i^*(S,\beta) 
&=&  \sum_{i=1}^N \Bigg[ (1-M_i)\Delta_i \log \Bigg( \sum_{j=1}^{V_i} D_{ij} (S_j)^{\exp({x}_i'\beta_X+z_i'\beta_Z)} \Bigg) + \nonumber \\ &&  (1-M_i)(1-\Delta_i) \log \Bigg( \sum_{j=V_i+1}^{J+1} D_{ij} (S_j)^{\exp({x}_i'\beta_X+z_i'\beta_Z)}  \Bigg) 
+ \nonumber \\ && M_i \log \Bigg( \sum_{j=1}^{J+1} D_{ij} (S_j)^{\exp({x}_i'\beta_X+z_i'\beta_Z)}  \Bigg)\Bigg],
\end{eqnarray}

\noindent where $\mathbf{\psi=[\beta,S]}$, $\beta=(\beta_X,\beta_Z)'$, and $\mathbf{S} = (S_1,S_2,...,S_{J+1})'$. Adopting the arguments from sections \ref{randsamptheory} and \ref{complexsurvtheory}, the weighted proposed estimator $\hat{\psi}_{\pi,X,Z}$ found by solving the weighted score equation $\sum_{i=1}^N \widecheck{U}_{i}({\psi}_{\pi,X,Z})=\sum_{i=1}^N \frac{1}{\pi_i} U_{i}({\psi}_{\pi,X,Z})=0$ can also be shown to be consistent for the true parameter $\hat{\psi}^0_{\pi}$ and asymptotically normal. These arguments only apply to settings in which the true covariate $X_i$ is used in the proposed method instead of $\hat{X}_i$. 

We will now make similar arguments of consistency and asymptotic normality for a new version of our log-likelihood that incorporates regression calibration. Begin by assuming that $\frac{n_C}{N} \rightarrow p\in (0,1)$, where $n_C$ is the number of subjects in the calibration subset. Recall that we assume the classical measurement error model, $X_i^{**}=X_i+\epsilon_i$, where $\epsilon_i \sim N(0,\sigma_{\epsilon_i}^2)$, as introduced in Section 2.3.1 from the main manuscript. Assume also that the following linear calibration model from the main manuscript holds: $X^{**}_i=\delta_{(0)}+\delta_{(1)}X_i^{*}+\delta_{(2)}Z_i+W_i$, where the random measurement error term $W_i \sim N(0,\sigma_{W_i}^2)$. Define  $\boldsymbol{\delta}=(\delta_{(0)},\delta_{(1)},\delta_{(2)})$. Then, since the vector of estimated nuisance parameters $\hat{\boldsymbol{\delta}}$ is a linear regression estimator, we can appeal to standard MLE theory to establish that it is consistent for the true vector of parameters $\boldsymbol{\delta}^0$ and asymptotically normal. To apply regression calibration, the first moment $\hat{X}_i=E(X_i|\boldsymbol{\delta},X_i^*,Z_i)$ is imputed in place of $X_i$ in our outcome model. To establish asymptotic normality of our regression calibration estimator, we first assume $\boldsymbol{\delta}$ is known and solve the following weighted log-likelihood equation $l_{\pi,X^*,Z}^*(\psi)=l(T_i,\Delta_i,Y_i^*,T_i^*,M_i,X^*_i, Z_i;\psi)$:

\begin{eqnarray}\label{loglikelihoodallNtheorysupp2}
l_{\pi,X^*,Z}^*(\psi) = \sum_{i=1}^N  \frac{1}{\pi_i}  l_i^*(S,\beta) 
&=&  \sum_{i=1}^N \Bigg[ (1-M_i)\Delta_i \log \Bigg( \sum_{j=1}^{V_i} D_{ij} (S_j)^{\exp(\hat{x}_i'\beta_X+z_i'\beta_Z)} \Bigg) + \nonumber \\ &&  (1-M_i)(1-\Delta_i) \log \Bigg( \sum_{j=V_i+1}^{J+1} D_{ij} (S_j)^{\exp(\hat{x}_i'\beta_X+z_i'\beta_Z)}  \Bigg) 
+ \nonumber \\ && M_i \log \Bigg( \sum_{j=1}^{J+1} D_{ij} (S_j)^{\exp(\hat{x}_i'\beta_X+z_i'\beta_Z)}  \Bigg)\Bigg],
\end{eqnarray}

\noindent where $\mathbf{\psi=[\beta,S]}$, $\beta=(\beta_X,\beta_Z)'$, and $\mathbf{S} = (S_1,S_2,...,S_{J+1})'$. We assume that distributions of the variables are such that when $X_i$ is replaced by $\hat{X}_i$, the log-likelihood $l_{\pi,X^*,Z}^*(\psi)=l(T_i,\Delta_i,Y_i^*,T_i^*,M_i,X^*_i, Z_i;\psi)$ remains continuously differentiable with respect to $\psi$ and that the Hessian matrix is still invertible. As before, we solve the weighted score equation $\sum_{i=1}^N \widecheck{U}_{i}^*({\psi}_{\pi,X^*,Z})=\sum_{i=1}^N \frac{1}{\pi_i} U_{i}^*({\psi}_{\pi,X^*,Z})=0$ in order to obtain our weighted proposed estimator, $\hat{\psi}^*_{\pi,X^*,Z}$. Under regularity conditions described previously, $\psi^*_{\pi}$ will be a unique, consistent solution to the vector of equations
$E\left[\frac{\partial l_{\pi}^*(\psi_{\pi,X^*,Z})}{\partial \psi_{\pi}}\right]=E\left[\sum_{i=1}^N \widecheck{U}_i^*({\psi}_{\pi,X^*,Z})\right]=0$. In general, $\psi^*_{\pi}$ is not the same as $\psi_{\pi}^0$, as the parameter estimates from regression calibration are viewed as an approximation \citep{buono10}. Using the techniques of  \citet{boos2013essential} once again, we can verify the asymptotic normality our estimator $\hat{\psi}^*_{\pi,X^*,Z}$, i.e.:

\begin{equation}\label{finaleqasymptnorm}
    \sqrt{N}(\hat{\psi}^*_{\pi,X^*,Z}-\psi^*_{\pi}) \xrightarrow{d} \mathcal{N}(0, {\text{V}}({\psi}^*_{\pi})).
\end{equation}

The regularity in equation \ref{finaleqasymptnorm} depends on known nuisance parameter vector $\boldsymbol{\delta}$ from the error model. With the additional usual regularity assumptions for linear regression that guarantee a consistent and asymptotically normal estimator for $\hat{\boldsymbol{\delta}}$, the regularity of our calibration estimator will still hold using this plug-in estimator for $\boldsymbol{\delta}$ by appealing to Theorem 5.31 in \citet{van2000asymptotic}. The variance ${\text{V}}({\psi}^*_{\pi})$ is estimated using the multiple imputation procedure introduced by \citet{baldoni2021use} described in the main text. When regression calibration is applied to the proposed method to adjust for covariate error, our estimator is only approximate but has been empirically shown to have minimal bias and good coverage probability when the true regression parameter is modest in size and the event of interest under study is rare.

\section{Supplemental details for HCHS/SOL data example}\label{HCHSSOLSUPPDETAILS}

We adopted the same exclusion criteria used in the ongoing clinical investigation that seeks to understand the relationship between several dietary intake variables and the risk of chronic diseases in the HCHS/SOL cohort. We excluded any participants who reported diabetes or unknown status at baseline ($N=3428$), had missing covariate data ($N=373$), or had no auxiliary follow-up ($N=551$), resulting in $12,317$ eligible participants. To mimic the planned analysis of the clinical investigation, for the $351$ subjects in the data who reported a positive diabetes status after one or more missed annual follow-up calls, we imputed that the event happened at the midpoint of the missed follow-up times. For most subjects with a missed call, subjects subsequently reported no diabetes diagnosis had occurred since the last call and so a negative disease status was imputed for the prior follow-up calls. The proposed method is applied to a subset of $8,200$ HCHS/SOL cohort participants, including all eligible SOLNAS subset participants ($N=420$) and HCHS/SOL participants from primary sampling units (PSUs) with 4 or fewer members ($N=282$). The remaining subset members were selected by taking a random sample of $7498$ participants that had not yet been selected.

\newpage

\clearpage
\begin{table}[ht]
\centering
      \begin{threeparttable}[t]
            \caption{Simulation results are shown for exponential failure times assuming the Cox proportional hazards model with $X\sim Gamma(0.2,1)$ and $\beta=\log(1.5)$ for (1) the grouped time survival approach that uses the true outcome data from all periodic visits and (2) the standard interval-censored approach that does not incorporate auxiliary data. The median percent (\%) bias, median standard errors (ASE), empirical median absolute deviation (MAD) and coverage probabilities (CP) are given for 1000 simulated data sets.}\label{tablesuppNotErrProne}
   
      \begin{tabular}{cccccccccccccc}
\hline
\multicolumn{3}{l}{} & \multicolumn{4}{c}{\multirow{ 2}{*}{Gold Standard Every Year}}  & \multicolumn{4}{c}{{Gold Standard Year 4 Only}}& \multicolumn{1}{c}{} \\ \multicolumn{3}{l}{} & \multicolumn{4}{c}{}  & \multicolumn{4}{c}{(No Auxiliary Data)}& \multicolumn{1}{c}{} \\ \cmidrule(r){4-7} \cmidrule{8-11}  $MR$\tnote{1}  & $CR$\tnote{2} & $N$\tnote{3}   & \% Bias & ASE & MAD & CP & \% Bias & ASE & MAD & CP &  \multicolumn{1}{c}{RE\tnote{4}}  \\

\hline
 0.0 & 0.9 & 1000 &  -2.015 & 0.158 & 0.151 & 0.953 & -1.402 & 0.160 & 0.155 & 0.951 & 1.019 \\ 
  
 & & 10,000 &    1.025 & 0.048 & 0.050 & 0.948 & 1.279 & 0.048 & 0.051 & 0.951 & 1.022 \\

& 0.7 & 1000 &   0.207 & 0.100 & 0.095 & 0.950 & 0.614 & 0.107 & 0.106 & 0.950 & 1.104 \\

 & & 10,000 &      0.466 & 0.031 & 0.031 & 0.942 & 0.398 & 0.033 & 0.034 & 0.947 & 1.137 \\

 & 0.5 & 1000 &   1.061 & 0.084 & 0.082 & 0.938 & 2.020 & 0.099 & 0.102 & 0.947 & 1.361 \\

 & & 10,000 &   0.503 & 0.026 & 0.026 & 0.954 & 0.382 & 0.031 & 0.034 & 0.951 & 1.370 \\

 0.2 & 0.9 & 1000 &    0.863 & 0.164 & 0.160 & 0.953 & -0.378 & 0.181 & 0.183 & 0.951 & 1.190 \\

 & & 10,000 &      1.493 & 0.049 & 0.052 & 0.945 & 0.769 & 0.054 & 0.055 & 0.952 & 1.197 \\

 &  0.7 & 1000 &   1.966 & 0.103 & 0.108 & 0.939 & 0.377 & 0.120 & 0.116 & 0.954 & 1.319 \\

 & &  10,000 &   1.524 & 0.032 & 0.035 & 0.936 & 0.332 & 0.037 & 0.038 & 0.946 & 1.336 \\

 &  0.5 & 1000 &     3.572 & 0.087 & 0.095 & 0.920 & 2.084 & 0.111 & 0.116 & 0.947 & 1.607 \\

 &   & 10,000 &     2.453 & 0.027 & 0.029 & 0.914 & 0.247 & 0.034 & 0.036 & 0.952 & 1.606 \\ 
  
  0.4 & 0.9 & 1000 &     1.935 & 0.176 & 0.181 & 0.944 & 1.178 & 0.213 & 0.222 & 0.959 & 1.411 \\

 & & 10,000 &    1.595 & 0.053 & 0.053 & 0.941 & 2.122 & 0.062 & 0.064 & 0.960 & 1.411 \\ 
  
 &  0.7 & 1000 &  2.100 & 0.110 & 0.117 & 0.927 & 1.616 & 0.140 & 0.138 & 0.958 & 1.586 \\

 & &  10,000 &     2.379 & 0.034 & 0.039 & 0.913 & 0.758 & 0.043 & 0.044 & 0.946 & 1.586 \\

 &  0.5 & 1000 &     6.148 & 0.093 & 0.106 & 0.911 & 3.186 & 0.130 & 0.136 & 0.952 & 1.911 \\

 &   & 10,000 &     4.344 & 0.029 & 0.030 & 0.885 & 0.122 & 0.040 & 0.043 & 0.945 & 1.893 \\ 
 \hline
\end{tabular}
 \begin{tablenotes}
    \item[1]  $MR=$ Average probability that the gold standard indicator $\Delta$ is missing at year 4 
      \item[2]  $CR=$ Average  censoring rate for the latent true event time at the end of study 
      \item[3]  $N=$ Sample size for proposed approach; if $MR > 0.0$, sample size for no auxiliary data approach is smaller because of missingness in gold standard indicator  $\Delta$.
       \item[4]  $RE=$ median relative efficiency, calculated as the median of the ratio of the estimated variance of the standard, no auxiliary data approach estimator to the estimated variance of the proposed method estimator , e.g. $\frac{Var({\hat{\beta}_{Standard}})}{Var(\hat{\beta}_{Proposed})}$
   \end{tablenotes}
    \end{threeparttable}
\end{table}

\begin{table}
\centering
      \begin{threeparttable}[t]
            \caption{Simulation results are shown for exponential failure times assuming the Cox proportional hazards model with $X\sim Gamma(0.2,1)$, $\beta=\log(1.5)$, and values of $Se = 0.90$ and $Sp = 0.80$ for the auxiliary data. The median percent (\%) bias, median standard errors (ASE), empirical median absolute deviation (MAD) and coverage probabilities (CP) are given for 1000 simulated data sets for the proposed method and the standard interval-censored approach that does not incorporate auxiliary data. }\label{supptableSESP}
   
      \begin{tabular}{cccccccccccccc}
\hline
\multicolumn{3}{l}{} & \multicolumn{4}{c}{Proposed}  & \multicolumn{4}{c}{No Auxiliary Data} & \multicolumn{1}{c}{ } \\ \cmidrule(r){4-7} \cmidrule{8-11}  $MR$\tnote{1}  & $CR$\tnote{2} & $N$\tnote{3}   & \% Bias & ASE & MAD & CP & \% Bias & ASE & MAD & CP & \multicolumn{1}{c}{RE\tnote{4}}  \\ 

\hline
 0.0 & 0.9 & 1000 &  -0.810 & 0.160 & 0.151 & 0.953 & -1.402 & 0.160 & 0.155 & 0.951 & 1.011 \\

 & & 10,000 &      1.501 & 0.048 & 0.049 & 0.954 & 1.279 & 0.048 & 0.051 & 0.951 & 1.007 \\

& 0.7 & 1000 &    0.914 & 0.104 & 0.099 & 0.952 & 0.614 & 0.107 & 0.106 & 0.950 & 1.043 \\

 & & 10,000 &     0.391 & 0.032 & 0.032 & 0.945 & 0.398 & 0.033 & 0.034 & 0.947 & 1.065 \\

 & 0.5 & 1000 &    1.598 & 0.091 & 0.091 & 0.943 & 2.020 & 0.099 & 0.102 & 0.947 & 1.172 \\

 & & 10,000 &       0.433 & 0.028 & 0.029 & 0.947 & 0.382 & 0.031 & 0.034 & 0.951 & 1.178 \\

 0.2 & 0.9 & 1000 &     -1.283 & 0.175 & 0.177 & 0.956 & -0.378 & 0.181 & 0.183 & 0.951 & 1.040 \\

 & & 10,000 &     0.970 & 0.053 & 0.052 & 0.952 & 0.769 & 0.054 & 0.055 & 0.952 & 1.050 \\

 &  0.7 & 1000 &       0.273 & 0.111 & 0.115 & 0.957 & 0.377 & 0.120 & 0.116 & 0.954 & 1.135 \\

 & &  10,000 &       0.604 & 0.034 & 0.034 & 0.947 & 0.332 & 0.037 & 0.038 & 0.946 & 1.152 \\ 
 
 &  0.5 & 1000 &      1.462 & 0.097 & 0.097 & 0.942 & 2.084 & 0.111 & 0.116 & 0.947 & 1.306 \\

 &   & 10,000 &    0.478 & 0.030 & 0.031 & 0.948 & 0.247 & 0.034 & 0.036 & 0.952 & 1.316 \\

 0.4 & 0.9 & 1000 &       -1.054 & 0.197 & 0.201 & 0.958 & 1.178 & 0.213 & 0.222 & 0.959 & 1.109 \\

 & & 10,000 &      1.388 & 0.059 & 0.059 & 0.953 & 2.122 & 0.062 & 0.064 & 0.960 & 1.127 \\

 &  0.7 & 1000 &        0.739 & 0.121 & 0.120 & 0.957 & 1.616 & 0.140 & 0.138 & 0.958 & 1.278 \\

 & &  10,000 &     0.762 & 0.037 & 0.038 & 0.952 & 0.758 & 0.043 & 0.044 & 0.946 & 1.306 \\

 &  0.5 & 1000 &      2.252 & 0.103 & 0.104 & 0.942 & 3.186 & 0.130 & 0.136 & 0.952 & 1.553 \\

 &   & 10,000 &        0.550 & 0.032 & 0.033 & 0.949 & 0.122 & 0.040 & 0.043 & 0.945 & 1.549 \\

  \hline

\end{tabular}
 \begin{tablenotes}
    \item[1]  $MR=$ Average probability that the gold standard indicator $\Delta$ is missing at year 4 
      \item[2]  $CR=$ Average  censoring rate for the latent true event time at the end of study 
      \item[3]  $N=$ Sample size for proposed approach; if $MR > 0.0$, sample size for no auxiliary data approach is smaller because of missingness in gold standard indicator  $\Delta$.
       \item[4]  $RE=$ median relative efficiency, calculated as the median of the ratio of the estimated variance of the standard, no auxiliary data approach estimator to the estimated variance of the proposed method estimator , e.g. $\frac{Var({\hat{\beta}_{Standard}})}{Var(\hat{\beta}_{Proposed})}$
   \end{tablenotes}
    \end{threeparttable}
\end{table}

\begin{table}
\centering
      \begin{threeparttable}[t]
            \caption{Simulation results are shown for exponential failure times assuming the Cox proportional hazards model with $X\sim Gamma(0.2,1)$ and $\beta=\log(3)$. The median percent (\%) bias, median standard errors (ASE), empirical median absolute deviation (MAD) and coverage probabilities (CP) are given for 1000 simulated data sets for the proposed method and the standard interval-censored approach that does not incorporate auxiliary data. Here, $Se = 0.80$ and $Sp = 0.90$ for the auxiliary data.}\label{supptablelog3}
   
      \begin{tabular}{cccccccccccccc}
\hline
\multicolumn{3}{l}{} & \multicolumn{4}{c}{Proposed}  & \multicolumn{4}{c}{No Auxiliary Data} & \multicolumn{1}{c}{ } \\ \cmidrule(r){4-7} \cmidrule{8-11}  $MR$\tnote{1}  & $CR$\tnote{2} & $N$\tnote{3}   & \% Bias & ASE & MAD & CP & \% Bias & ASE & MAD & CP & \multicolumn{1}{c}{RE\tnote{4}}  \\ 

\hline
 0.0 & 0.9 & 1000 &   0.627 & 0.126 & 0.121 & 0.961 & 0.937 & 0.135 & 0.136 & 0.950 & 1.155 \\

 & & 10,000 &        0.032 & 0.039 & 0.040 & 0.940 & 0.150 & 0.042 & 0.042 & 0.939 & 1.162 \\

& 0.7 & 1000 &      1.082 & 0.114 & 0.118 & 0.953 & 0.657 & 0.130 & 0.134 & 0.949 & 1.274 \\

 & & 10,000 &     0.207 & 0.036 & 0.035 & 0.946 & 0.301 & 0.041 & 0.043 & 0.955 & 1.281 \\

 & 0.5 & 1000 &      0.393 & 0.122 & 0.120 & 0.951 & 0.517 & 0.147 & 0.151 & 0.952 & 1.474 \\

 & & 10,000 &      0.302 & 0.038 & 0.038 & 0.954 & 0.164 & 0.046 & 0.046 & 0.950 & 1.458 \\

 0.2 & 0.9 & 1000 &    1.088 & 0.132 & 0.130 & 0.955 & 1.493 & 0.151 & 0.149 & 0.955 & 1.286 \\

 & & 10,000 &     0.175 & 0.041 & 0.042 & 0.939 & 0.124 & 0.047 & 0.046 & 0.944 & 1.298 \\

 &  0.7 & 1000 &      1.243 & 0.119 & 0.118 & 0.955 & 1.213 & 0.145 & 0.149 & 0.950 & 1.464 \\

 & &  10,000 &    0.285 & 0.037 & 0.038 & 0.951 & 0.436 & 0.045 & 0.048 & 0.955 & 1.473 \\

 &  0.5 & 1000 &     0.505 & 0.126 & 0.124 & 0.950 & 0.502 & 0.165 & 0.179 & 0.939 & 1.724 \\

 &   & 10,000 &     0.165 & 0.040 & 0.039 & 0.950 & 0.041 & 0.052 & 0.051 & 0.948 & 1.708 \\

 0.4 & 0.9 & 1000 &      1.011 & 0.141 & 0.140 & 0.949 & 1.840 & 0.175 & 0.184 & 0.952 & 1.515 \\

 & & 10,000 &     0.049 & 0.044 & 0.044 & 0.950 & 0.084 & 0.054 & 0.053 & 0.954 & 1.522 \\

 &  0.7 & 1000 &      1.275 & 0.125 & 0.125 & 0.948 & 1.897 & 0.168 & 0.180 & 0.944 & 1.789 \\

 & &  10,000 &        0.115 & 0.039 & 0.041 & 0.958 & 0.509 & 0.052 & 0.053 & 0.956 & 1.792 \\

 &  0.5 & 1000 &      0.692 & 0.130 & 0.134 & 0.944 & 2.561 & 0.193 & 0.205 & 0.942 & 2.174 \\

 &   & 10,000 &      0.215 & 0.041 & 0.042 & 0.948 & -0.140 & 0.060 & 0.058 & 0.949 & 2.118 \\

  \hline

\end{tabular}
 \begin{tablenotes}
\item[1]  $MR=$ Average probability that the gold standard indicator $\Delta$ is missing at year 4 
      \item[2]  $CR=$ Average  censoring rate for the latent true event time at the end of study 
      \item[3]  $N=$ Sample size for proposed approach; if $MR > 0.0$, sample size for no auxiliary data approach is smaller because of missingness in gold standard indicator  $\Delta$.
       \item[4]  $RE=$ median relative efficiency, calculated as the median of the ratio of the estimated variance of the standard, no auxiliary data approach estimator to the estimated variance of the proposed method estimator , e.g. $\frac{Var({\hat{\beta}_{Standard}})}{Var(\hat{\beta}_{Proposed})}$
   \end{tablenotes}
    \end{threeparttable}
\end{table}

\begin{table}
\centering
      \begin{threeparttable}[t]
            \caption{Simulation results are shown for data simulated to be from a complex survey with exponential failure times assuming the Cox proportional hazards model with $X\sim Normal(\textrm{shape}_s+\omega_{gs},\textrm{scale}_s+\rho_{gs})$ for an individual in block group $g$ and stratum $s$ and $\beta=\log(1.5)$. The median percent (\%) bias, median standard errors (ASE), median absolute deviation (MAD) and coverage probabilities (CP) are given for 1000 simulated data sets for the weighted proposed estimator and the weighted interval-censored approach that does not incorporate auxiliary data when both use a sandwich variance estimator to address within-cluster correlation. Here, $Se = 0.80$ and $Sp = 0.90$ for the auxiliary data.}\label{weightedtablenormalMED}
   
      \begin{tabular}{cccccccccccccc}
\hline
\multicolumn{3}{l}{} & \multicolumn{4}{c}{Proposed}  & \multicolumn{4}{c}{No Auxiliary Data} & \multicolumn{1}{c}{ } \\ \cmidrule(r){4-7} \cmidrule{8-11}  $MR$\tnote{1}  & $CR$\tnote{2} & $N$\tnote{3}   & \% Bias & ASE & MAD & CP & \% Bias & ASE & MAD & CP & \multicolumn{1}{c}{RE\tnote{4}}  \\ 

\hline
 0.0 & 0.9 & 1000 & -4.213 & 0.122 & 0.139 & 0.933 & -4.642 & 0.124 & 0.138 & 0.938 & 1.018 \\

 & & 10,000 &     -3.293 & 0.042 & 0.041 & 0.933 & -3.285 & 0.042 & 0.041 & 0.936 & 1.010 \\

& 0.7 & 1000 &     -0.778 & 0.078 & 0.081 & 0.936 & -0.983 & 0.079 & 0.080 & 0.932 & 1.024 \\

 & & 10,000 &      -0.239 & 0.026 & 0.028 & 0.937 & -0.331 & 0.026 & 0.028 & 0.937 & 1.029 \\

 & 0.5 & 1000 &    -0.459 & 0.062 & 0.066 & 0.936 & -0.516 & 0.065 & 0.067 & 0.932 & 1.066 \\

 & & 10,000 &  -0.210 & 0.020 & 0.021 & 0.938 & -0.225 & 0.021 & 0.022 & 0.941 & 1.087 \\

 0.2 & 0.9 & 1000 &    -3.237 & 0.133 & 0.152 & 0.926 & -3.294 & 0.137 & 0.157 & 0.937 & 1.055 \\

 & & 10,000 &   -3.275 & 0.046 & 0.045 & 0.942 & -3.160 & 0.047 & 0.047 & 0.941 & 1.058 \\

 &  0.7 & 1000 &   -1.547 & 0.083 & 0.085 & 0.929 & -1.153 & 0.088 & 0.094 & 0.931 & 1.117 \\

 & &  10,000 &     -0.530 & 0.027 & 0.028 & 0.939 & -0.194 & 0.029 & 0.029 & 0.937 & 1.124 \\

 &  0.5 & 1000 &   -0.194 & 0.066 & 0.070 & 0.932 & -0.558 & 0.072 & 0.076 & 0.930 & 1.190 \\

 &   & 10,000 &     -0.143 & 0.021 & 0.022 & 0.946 & -0.086 & 0.024 & 0.024 & 0.941 & 1.230 \\

 0.4 & 0.9 & 1000 &    -2.592 & 0.149 & 0.167 & 0.924 & -2.799 & 0.155 & 0.165 & 0.930 & 1.108 \\

 & & 10,000 &   -2.977 & 0.051 & 0.050 & 0.934 & -2.789 & 0.054 & 0.054 & 0.932 & 1.131 \\

 &  0.7 & 1000 &   -0.665 & 0.090 & 0.091 & 0.930 & -0.173 & 0.101 & 0.108 & 0.933 & 1.266 \\

 & &  10,000 &  -0.564 & 0.029 & 0.031 & 0.939 & -0.337 & 0.033 & 0.035 & 0.947 & 1.286 \\

 &  0.5 & 1000 &   0.471 & 0.070 & 0.072 & 0.926 & 0.081 & 0.082 & 0.089 & 0.920 & 1.407 \\ 

 &   & 10,000 &    -0.193 & 0.023 & 0.024 & 0.947 & -0.063 & 0.027 & 0.029 & 0.941 & 1.455 \\

  \hline
\end{tabular}
 \begin{tablenotes}
     \item[1]  $MR=$ Average probability that the gold standard indicator $\Delta$ is missing at year 4 
      \item[2]  $CR=$ Average  censoring rate for the latent true event time at the end of study
       \item[3]  $(N)=$ Average sample size for proposed approach; if $MR > 0.0$, sample size for no auxiliary data approach is smaller because of missingness in gold standard indicator  $\Delta$.
       \item[4]  $RE=$ median relative efficiency, calculated as the median of the ratio of the estimated variance of the standard, no auxiliary data approach estimator to the estimated variance of the proposed method estimator , e.g. $\frac{Var({\hat{\beta}_{Standard}})}{Var(\hat{\beta}_{Proposed})}$
   \end{tablenotes}
    \end{threeparttable}
\end{table}

\begin{table}
\centering
 \begin{threeparttable}[t]
\caption {Sensitivity analysis using HCHS/SOL data on a subset of study participants with visit 2 sensitivity ($Se = 0.77$) and specificity ($Sp = 0.92$) values. Hazard Ratio (HR) and 95\% confidence interval (CI) estimates of incident diabetes for a 20\% increase in consumption of energy (kcal/d), protein (g/d), and protein density (\% energy from protein/d) based on the proposed estimator and the interval-censored approach that does not incorporate auxiliary data.} \label{tableHRSensAnalysis} 

\centering
\begin{tabular}{llll}
\hline
\multicolumn{1}{c}{}  & \multicolumn{2}{c}{HR (95\% CI)} & \multicolumn{1}{c}{ } \\ \cmidrule(r){2-3}  \multicolumn{1}{c}{Model\tnote{1}} & \multicolumn{1}{c}{Proposed} & \multicolumn{1}{c}{No Auxiliary Data}  & \multicolumn{1}{c}{RE\tnote{2}} \\  
  \hline
 Energy (kcal/d)   &    1.26 (0.49, 3.24) & 1.20 (0.41, 3.82) & 1.27 \\

   Protein (g/d)   &    1.37 (0.88, 2.14) & 1.37 (0.74, 2.51) & 1.85 \\

  Protein Density   &       1.01 (1.00, 1.03) & 1.01 (1.00, 1.03) & 1.43 \\

   \hline
\end{tabular}
 \begin{tablenotes}
     \item[1] Each model is adjusted for potential confounders including age, body mass index (BMI), sex, Hispanic/Latino background, language preference, education, income, and smoking status.
           \item[2]  $RE=$ relative efficiency, calculated as the ratio of the estimated variance of the standard, no auxiliary data approach estimator to the estimated variance of the proposed method estimator , e.g. $\frac{Var({\hat{\beta}_{Standard}})}{Var(\hat{\beta}_{Proposed})}$

   \end{tablenotes}
    \end{threeparttable}%
\end{table}

\clearpage

\bibliography{biblio.bib}

\end{document}